\documentclass[]{raa}            
\usepackage{graphicx,times,epsf}             
\usepackage{amsmath}
\usepackage{makecell}
\usepackage[utf8]{inputenc}
\usepackage{multirow}
\usepackage{makecell}
\usepackage{adjustbox}

\newcommand{\kms}{km\,s$^{-1}$}

\begin{document}
   \title{ALMA observations of the circumstellar envelope around EP Aqr}
   \volnopage{Vol.0 (200x) No.0, 000--000}      
   \setcounter{page}{1}          
   \author{P.T. Nhung\inst{1,2}
     \and
     D.T. Hoai\inst{1,2,4}
     \and
     P. Tuan-Anh\inst{1}
     \and
     T. Le\,Bertre\inst{2}
     \and
     P. Darriulat\inst{1}
     \and
     P.N. Diep\inst{1}
     \and N.T. Phuong\inst{1}
     \and N.T. Thao\inst{1}
     \and J.M. Winters\inst{3}\\}    
          
   \institute{Department of Astrophysics, Vietnam National Space Center, VAST, 18 Hoang Quoc Viet, Cau Giay, Ha Noi, Vietnam; {\it pttnhung@vnsc.org.vn}
   \and LERMA, UMR 8112, CNRS and Observatoire de Paris, PSL Research University,
	      61 av. de l'Observatoire, F-75014 Paris, France
         \and IRAM, 300 rue de la Piscine, Domaine Universitaire, 
	 F-38406 St. Martin d'H\`eres, France
   \and Graduate University of Science and Technology, VAST, 18 Hoang Quoc Viet, Cau Giay, Ha Noi, Vietnam} 
   
   \date{}
   \abstract {Atacama Large Millimetre/sub-millimetre Array (ALMA) observations of the CO(1-0) and CO(2-1) emissions from the circumstellar
envelope of the Asymptotic Giant Branch (AGB) star EP Aqr have been made with four times better spatial resolution than previously
available. They are analysed with emphasis on the de-projection in space of the effective emissivity and flux of matter using as
input a prescribed configuration of the velocity field, assumed to be radial. The data are found to display an intrinsic axi-symmetry
with respect to an axis making a  small angle with the line of sight. A broad range of wind configurations, from prolate (bipolar) to oblate (equatorial) has been studied and found to be accompanied by significant equatorial emission. Qualitatively, the effective emissivity is enhanced near the equator to produce the central narrow
component observed in the Doppler velocity spectra and its dependence on star latitude generally follows that of the wind velocity
with the exception of an omni-present depression near the poles. In particular, large equatorial expansion velocities produce a flared
disc or a ring of effective emissivity and mass loss. The effect on the determination of the orientation of the star axis of
radial velocity gradients and possibly competing rotation and expansion in the equatorial disc is discussed. In general,
the flux of matter is found to reach a broad maximum at distances of the order of 500 au from the star. Arguments are given that
may be used to prefer one wind velocity distribution to
another. As a result of the improved quality of the data, a deeper understanding of the constraints imposed on morphology and
kinematics has been obtained.  
   }
   
   \titlerunning{ALMA observations of EP Aqr}
   \authorrunning{P.T. Nhung, D.T. Hoai, P. Tuan-Anh et al.}
  
   \maketitle

%

   \section{Introduction}
   EP Aqr is an oxygen-rich M type semi-regular variable AGB star at a distance
of only 114 $\pm$ 8 pc from the solar system (van Leeuwen 2007), making it one
of the best studied such stars. While several observations suggest that it
entered recently the AGB and is at the beginning of its evolution on the
Thermally Pulsing phase, others point to a longer mass loss episode. Among
the former are the luminosity (3450 solar luminosities), the absence of
Technetium in the spectrum (Lebzelter and Hron 1999) and the low value
($\sim$10) of the $^{12}$C to $^{13}$C abundance ratio (Cami et al. 2000).
Among the latter are the Herschel 70 $\mu$m observation (Cox et al. 2012)
of a large trailing circumstellar shell and the HI observation
(Le Bertre and G\'erard 2004) of its interaction with the interstellar
medium, together suggesting a mass loss episode at the scale of
10$^4$ to 10$^5$ years.

EP Aqr belongs to the class of rare AGB stars that show a
composite CO line-profile (Knapp et al. 1998). A narrow component 
(FWHM $\sim$ 2-5 \kms) is overimposed on a broader one (FWHM $\sim$ 
20 \kms). Interferometric data (e.g. on RS Cnc : Libert et
al. 2010, Hoai et al. 2014) have shown that these sources exhibit an
axi-symmetric structure with a bipolar outflow, origin of the broad 
component, and an equatorial structure, origin of the narrow
component. Position-velocity diagrams discussed by Libert et
al. (2010) indicate that the equatorial structure (on scale of 100 -
1000 au) is in expansion rather than in rotation. Bipolar structures 
may also be present in AGB stars that do not show a composite
line-profile in CO (e.g. RX Boo, Castro-Carrizo et al. 2010),
suggesting that bipolarity might not be so rare among AGB stars.

Recently, observations of \mbox{$^{12}$CO(1-0)} and \mbox{$^{12}$CO(2-1)} emissions
using the IRAM 30-m telescope and the Plateau-de-Bure Interferometer
have been reported (Winters et al. 2003, 2007 and Nhung et al. 2015a)
and have been shown to display features characteristic of both a radial
dependence of the flux of matter and the presence of a bipolar outflow.
Here and in the remainder of the paper, we use the expression
  ``flux of matter'' to emphasize the dependence on direction of the mass loss process.

The sky maps of the observed intensity display approximate circular
symmetry about the central star, suggesting a circumstellar envelope having
morphology either spherically symmetric or axi-symmetric with its axis nearly
parallel to the line of sight. As rotation about such an axis would not be
detectable, and as the sky map of the mean Doppler velocity does not display
significant asymmetry, it is natural to assume that the wind is essentially
radial, as typically expected from the expanding circumstellar envelope of
mass-losing stars in the early AGB phase. Under such a hypothesis, it is in
principle possible to assume any form for the wind velocity field and deduce
from it the distribution of the effective emissivity in space. The only
requirement is for the radial expansion velocity to be large enough to project
on the line of sight as the largest Doppler velocities that are observed.
This feature was discussed in some detail in Nhung et al. (2015a) and
Diep et al. (2016) and is again extensively exploited in the present article,
which analyses new ALMA observations of \mbox{$^{12}$CO(1-0)} and \mbox{$^{12}$CO(2-1)}
emissions having four times better spatial resolution than the preceding
IRAM observations. This substantial improvement makes it possible to study,
more reliably than before, the possible presence of irregularities in the
morphology of the circumstellar envelope, causing it to depart from exact
axi-symmetry. In the wake of an argument by Knapp et al. (1998), such features
have been suggested by earlier analyses (Nakashima 2006, Winters et al. 2007)
as possibly associated with an episode of increased mass loss rate. While the
interpretation of the IRAM observations favours a bipolar outflow over a
spherical wind with a strong radial modulation (Le Bertre et al. 2016),
both may to some extent co-exist: an aim of the present article is to
clarify this issue.

The paper is organised as follows: after having briefly described
in Section 2 the conditions under which the observations were made and
having summarized the main points relating to data reduction, we explain
in Section 3 the details of the method of de-projection used in the present
analysis to produce a 3-D distribution of the effective emissivity and we
present the results. Section 4 comments on possible interpretations.
Summary and conclusions are presented in Section 5.

\section{Observations and data reduction}

\begin{figure*}
\centering
\includegraphics[width=0.45\textwidth,trim=0cm .8cm 0.cm 1.3cm,clip]{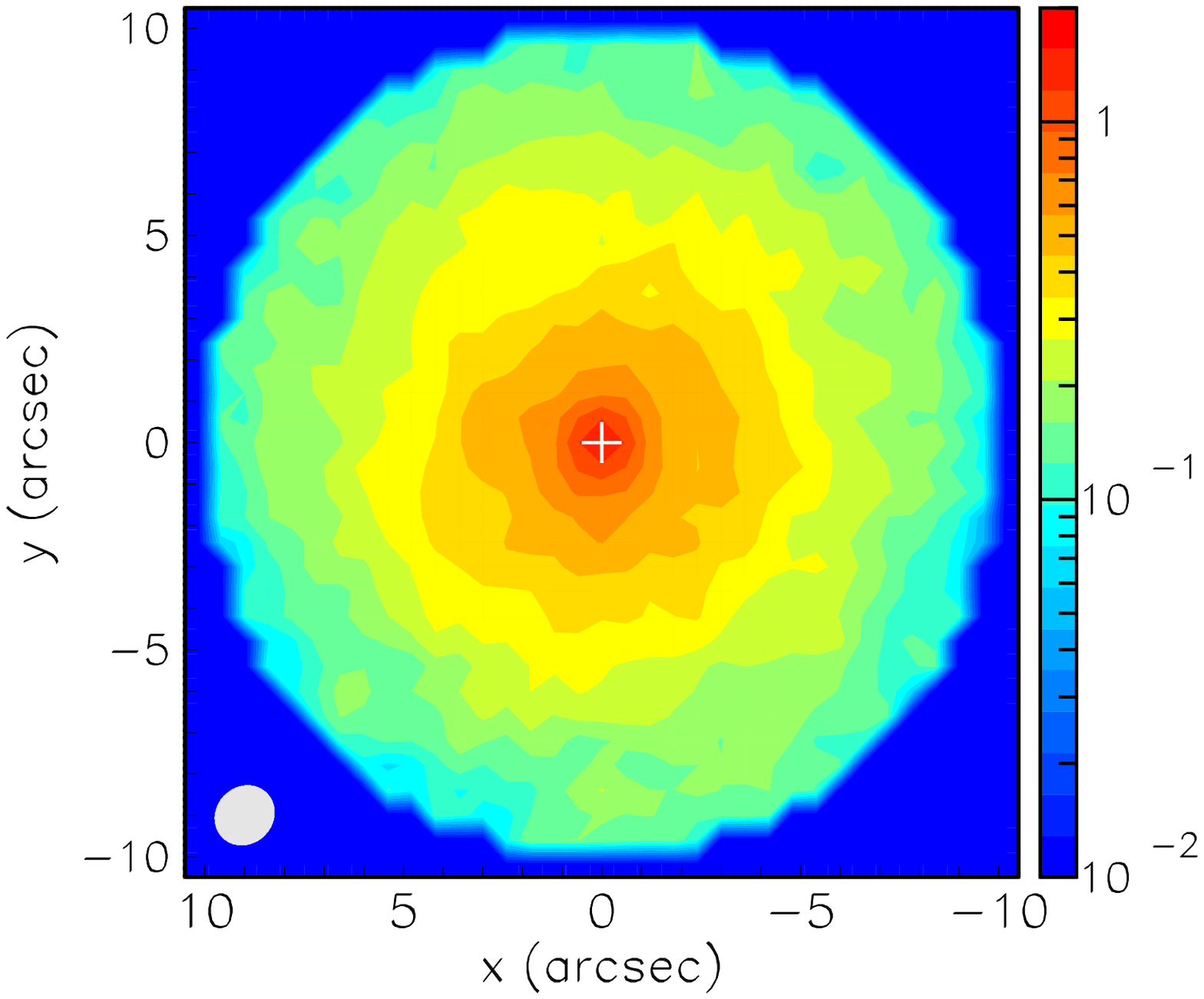}
\includegraphics[width=0.45\textwidth,trim=0cm .8cm 0.cm 1.3cm,clip]{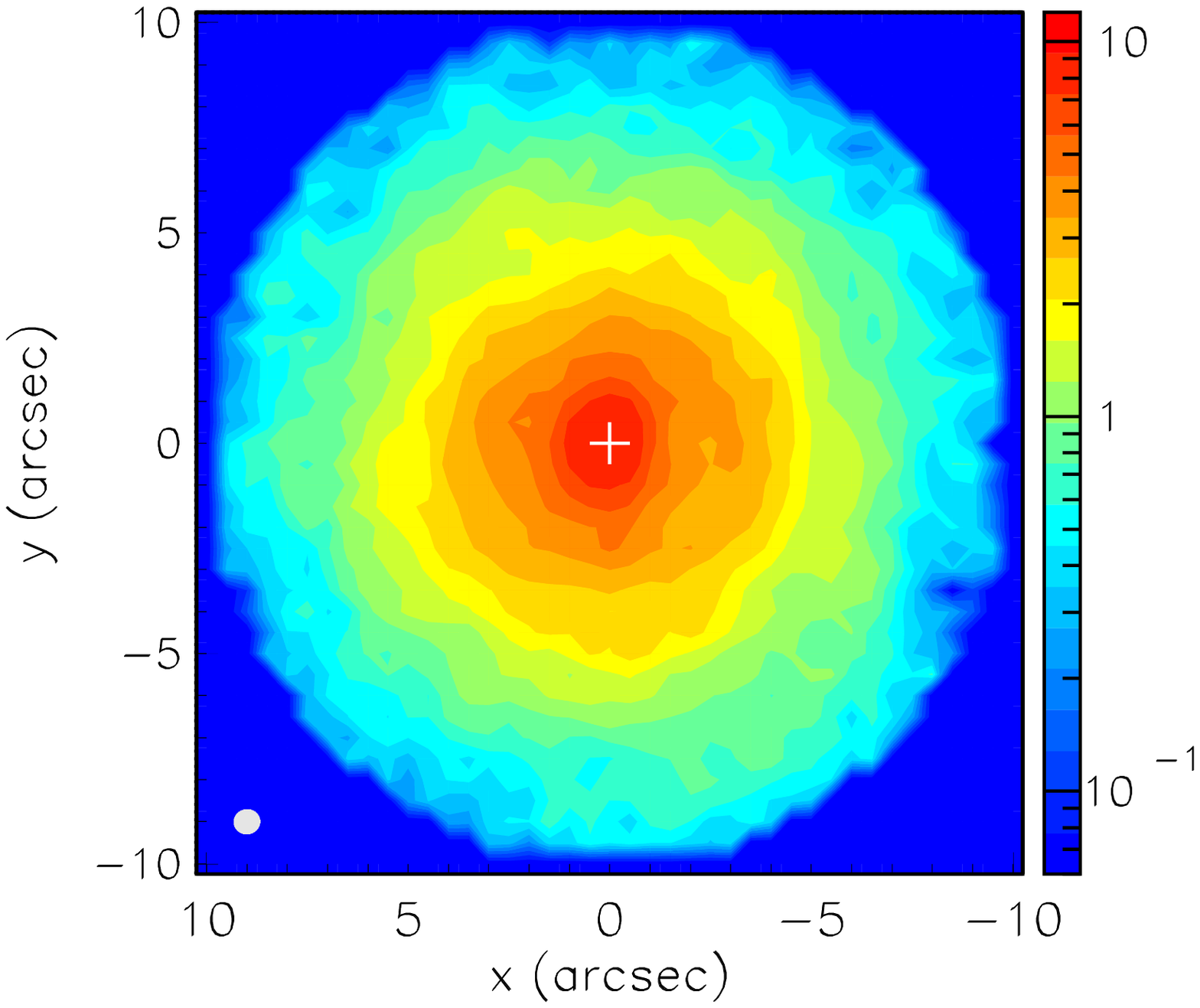}
\includegraphics[width=0.45\textwidth,trim=0cm .8cm 0.cm 1.3cm,clip]{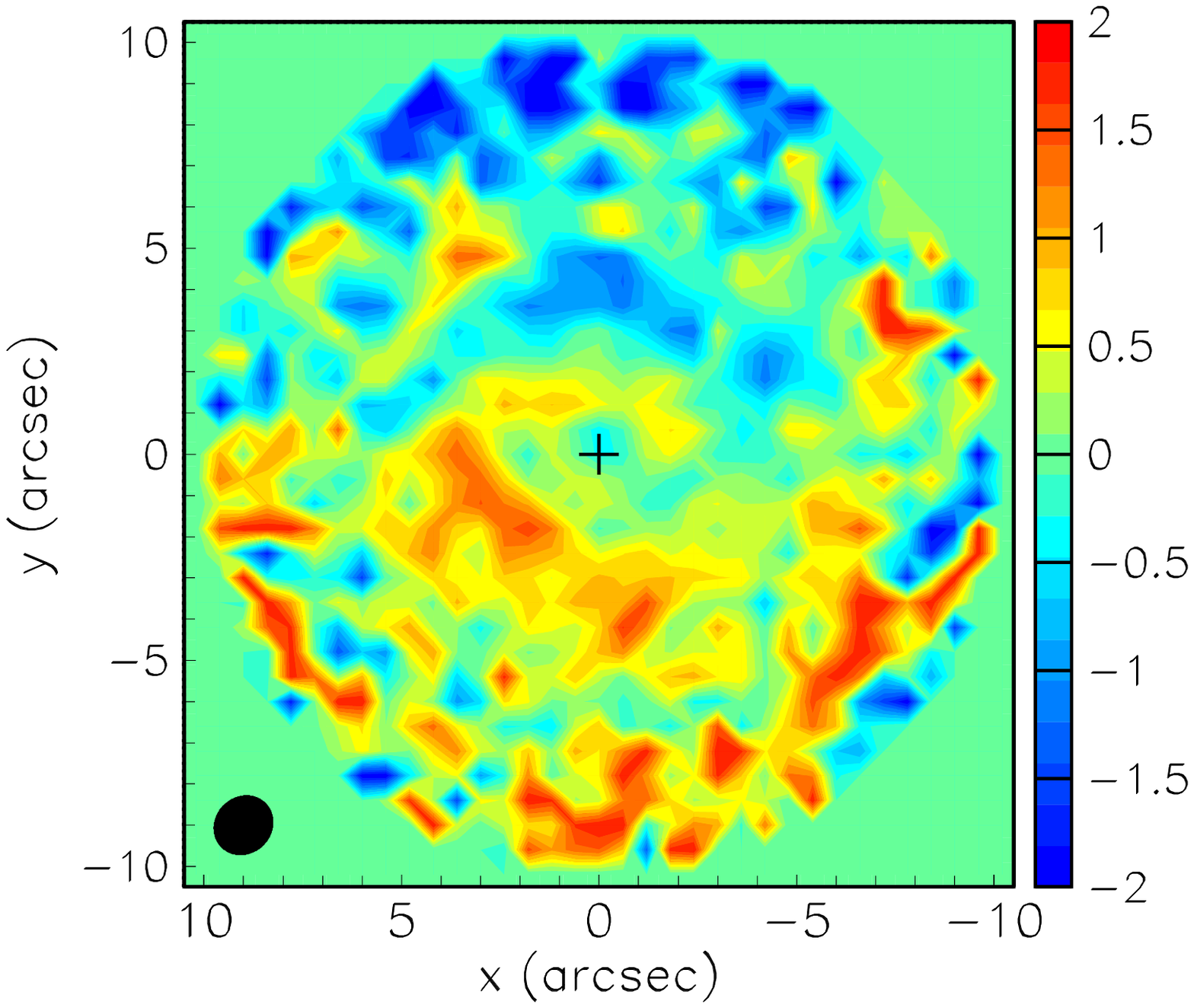}
\includegraphics[width=0.45\textwidth,trim=0cm .8cm 0.cm 1.3cm,clip]{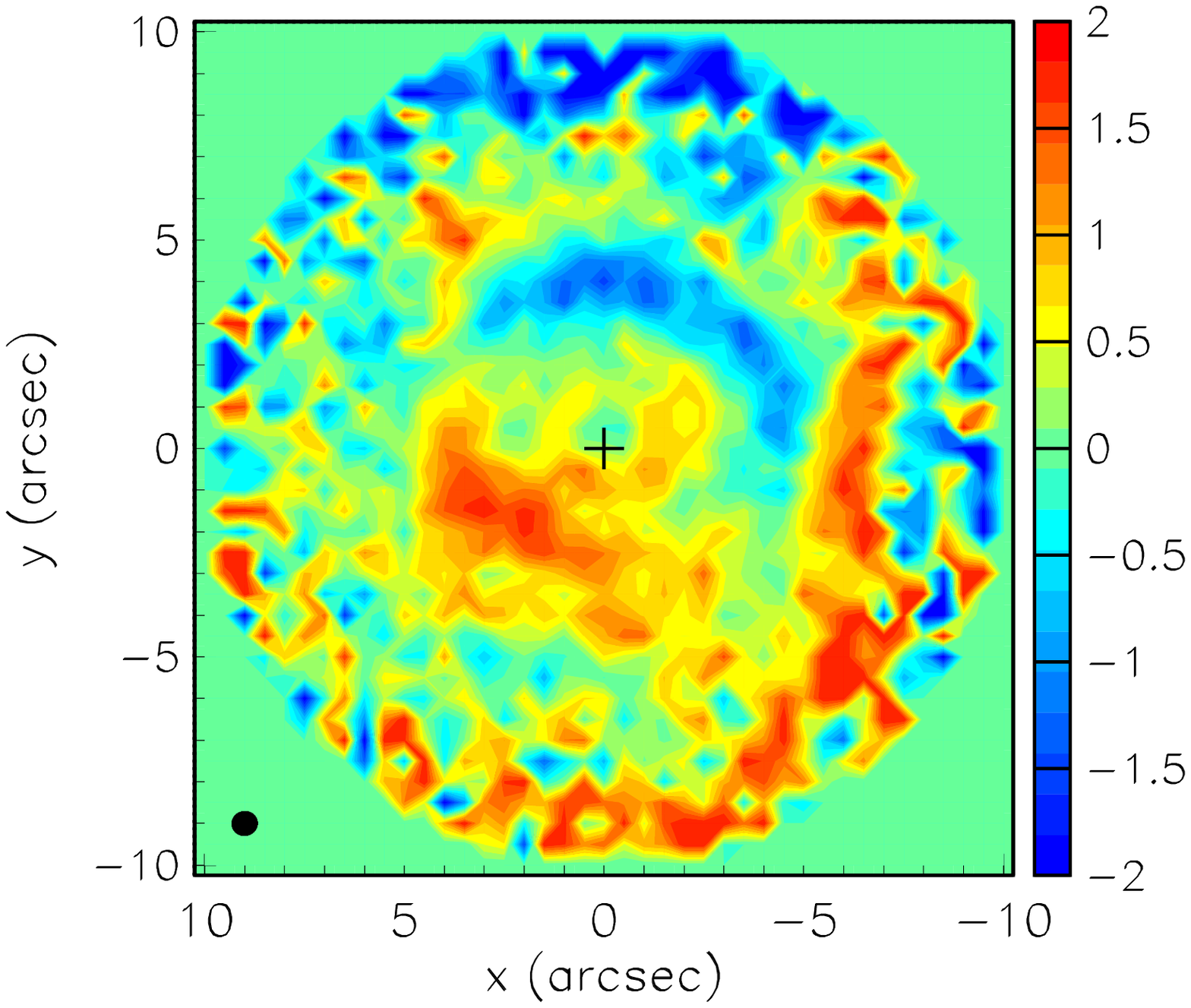}
\caption{Sky maps of the integrated intensity (upper panels) and mean Doppler velocity (lower panels) for CO(1-0) (left) and CO(2-1) (right) emission. The intensity maps are integrated between $-$12 and 12 \kms\ with respect to the stellar $V_{lsr}$. Units are Jy arcsec$^{-1}$\kms\ and \kms. The beams are shown in the lower left corners.}
\label{fig1}
\end{figure*}

\begin{figure*}
\centering
\includegraphics[width=0.9\textwidth,trim=0cm .8cm 0.cm 0.cm,clip]{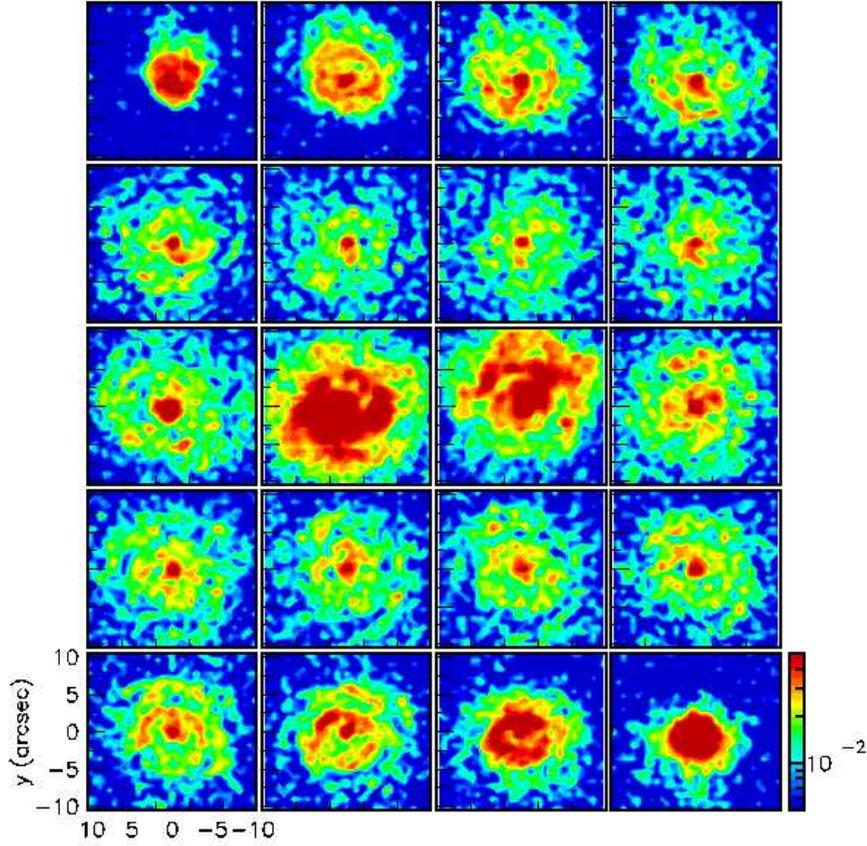}
\caption{Channel maps of \mbox{CO(1-0)} emission grouped in bins of 1 \kms\ between $-10$ \kms\ and 10 \kms\ with respect to the stellar $V_{lsr}$. The colour scale is in units of Jy\,arcsec$^{-2}$\,\kms.}
\label{fig2}
\end{figure*}

\begin{figure*}
\centering
\includegraphics[width=0.9\textwidth,trim=0cm .8cm 0.cm 0.cm,clip]{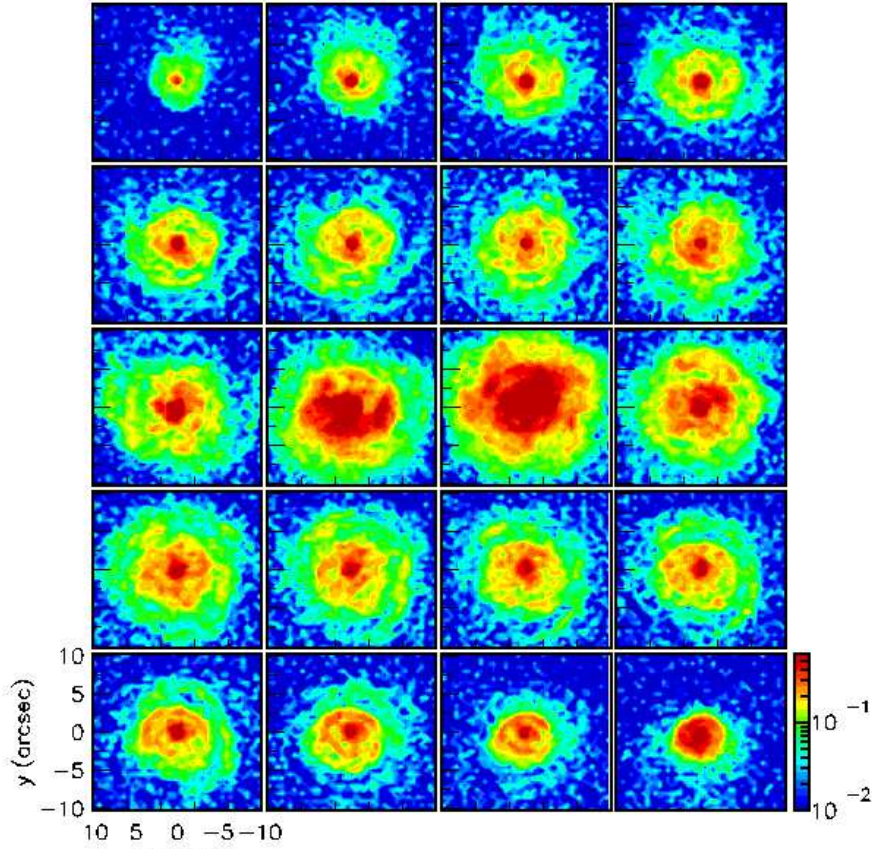}
\caption{Channel maps of \mbox{CO(2-1)} emission grouped in bins of 1 \kms\ between $-10$ \kms\ and 10 \kms\ with respect to the stellar $V_{lsr}$. The colour scale is in units of Jy\,arcsec$^{-2}$\,\kms.}
\label{fig3}
\end{figure*}

The observations used in the present article were made in cycle 4 of ALMA
operation (2016.1.00026.S) between October 30th 2016 and April 5th 2017.
\mbox{CO(1-0)} emission was observed in 3 execution blocks in mosaic mode
(3 pointings) with the number of antennas varying between 41 and 45.
\mbox{CO(2-1)} emission was observed in 2 execution blocks in mosaic mode
(10 pointings) with the number of antennas varying between 38 and 40.
The area covered by the mosaic is $\sim$60 arcsec north-south and $\sim$45 arcsec east-west.
Both lines were observed in two different configurations, C40-2 and
C40-5 and the data were then merged in the uv plane.
The correlator was set up with channel spacings of 30 and 60 kHz with widths of 3840 and 1920 channels for \mbox{CO(1-0)} and \mbox{CO(2-1)} respectively,
namely $\sim$80 m\,s$^{-1}$ in velocity resolution for both lines. We used the calibration scripts provided
by ALMA and the CASA 4.7.2 software package to obtain the calibrated data. Flux calibrators were
quasar J2148+0657 and Neptune. The calibrated data were then regridded in velocity to the LSR frame
and exported through UVFITS format to the GILDAS package for imaging. We use natural weighting
resulting in beam sizes which are given in Table~\ref{table1} together with other parameters of relevance to these observations.
The \mbox{CO(1-0)} and \mbox{CO(2-1)} data have been calibrated using CASA\footnote
{ http://casa.nrao.edu} and the mapping was done with GILDAS\footnote
{https://www.iram.fr/IRAMFR/GILDAS}. The origin of coordinates at
RA=21h 46m 31.848 s and DEC=$-$02$^\circ$12$'$ 45.93$''$ corresponds
to year 2000. Between 2000 and the time of observation, the source has
moved by 0.40 arcsec East and 0.31 arcsec North (proper motion of
[24.98, 19.54] mas/yr taken from van Leeuwen 2007); the data
have been corrected accordingly. The spectral resolution (channel)
has been smoothed to 0.2 \kms\ and we present results in a velocity range between
$-$20 and 20 \kms\ with respect to the LSR velocity of the source. We use as origin the Doppler velocity of $-$33.6 \kms\
(LSR) about which the profile is well symmetric. Channel maps and sky
maps of the integrated intensity and mean Doppler velocity are shown in Figures~\ref{fig1} to~\ref{fig3}.
Arcs and possibly spiral structure can be seen in these channel maps. However in the present paper we
  limit our study to the general morphology and kinematics. We will discuss the sub-structures in a forthcoming paper.

\begin{table}
\centering
\caption{Observing conditions.}
\begin{tabular}{|c|c|c|}
\hline
&CO(1-0)            &CO(2-1)           \\
\hline
Beam FWHM                    &$0.78''\times0.70''$ &$0.33''\times0.30''$ \\
\hline
Beam PA                      &PA=$-56^\circ$     &PA=$-80^\circ$    \\
\hline
Max. baseline                &1124 m             &1400 m            \\
\hline
Min. baseline                  &19 m               &15 m              \\
\hline
Time on source               &98.3 min           &52.4 min          \\
\hline
Zero spacing                 &No                 &No                \\
\hline
\makecell{Noise \\(mJy\,beam$^{-1}$/0.2 \kms\ bin)} &7                  &6                 \\
\hline
\makecell{Total flux, $R<13''$\\ (Jy\,\kms)} &110                &549               \\
\hline
\end{tabular}\\
\label{table1}
\end{table}

The aim of the present article is limited to a presentation of
the morphology and kinematics of the circumstellar envelope of EP Aqr
inferred directly from the observational data without invoking a physical
model. In addition to being a necessary preliminary to reliable physical modelling,
  it illustrates the type of constraints imposed on possible models.
A discussion based on such a physical model, accounting in particular
for the effect of temperature and absorption, is deferred to a forthcoming
paper, together with the presentation and analysis of data obtained from
other emission lines.

\section{ De-projection, effective emissivity and flux of matter}
\subsection{The method}
We use Cartesian coordinates centred on the star with the $x$ axis pointing
east, the $y$ axis pointing north and the $z$ axis parallel to the line of
sight and pointing away from us. We also occasionally use polar coordinates
in the plane of the sky, with radius $R=\sqrt{x^2+y^2}$ and position angle
$\psi=\tan^{-1}(y/x)-90^\circ$ measured clockwise from north. We generally
assume radial winds having axi-symmetric velocities about an axis making
an angle $\varphi$ with the line of sight and projecting on the plane of the
sky at position angle $\theta$. The projected axis is used as origin of
stellar longitude $\omega$ and we call $\alpha$ the stellar latitude. The
distance in space from the star is $r=\sqrt{R^2+z^2}$. Finally we call
\textit{\textbf{V}} the velocity vector and $V_z$ the Doppler velocity, projection of \textit{\textbf{V}}
on the $z$ axis.

Transformation relations from sky to stellar coordinates $(x',y',z')$ read:
\begin{equation}
  \begin{split}
	x'&= x \cos\theta + y \sin\theta\\
	y'&=-x \sin\theta \cos\varphi +y \cos\theta \cos\varphi +z \sin\varphi\\
	z'&= x \sin\theta \sin\varphi -y \cos\theta \sin\varphi +z \cos\varphi\\
  \end{split}
\end{equation}

Observations provided by a radio telescope are in the form of data-cube
elements, $f(x,y,V_z)$ giving for each point $(x,y)$ on the plane of the
sky (pixel) the distribution of the measured brightness as a function of
Doppler velocity $V_z$. Before attempting to give a physical interpretation
of what is observed, it is necessary to apply a de-projection to the observed
image, namely to calculate, for each space point $(x,y,z)$ the effective
emissivity $\rho$ and the three components of the velocity-vector
(Diep et al. 2016). This is a largely under-determined problem and, in
general, it cannot be solved. However, it becomes solvable once we know
the velocity field $(V_x,V_y,V_z)$ at each point in space. In that case,
for each pixel $(x,y)$ the dependence on $z$ of the Doppler velocity is
known and, to the extent that the corresponding relation can be inverted,
it is possible to associate to each Doppler velocity $V_z$ a point $(x,y,z)$
in space. From the definition of the effective emissivity $\rho$, the intensity
measured in this pixel reads
\begin{equation}
  F(x,y)=\int{f(x,y,V_z)dV_z}=\int{\rho(x,y,z)dz}
\end{equation}
where the first integral spans the Doppler velocity range covered
by the observation and the second integral extends over the whole line of
sight, implying
\begin{equation}
  \rho(x,y,z)=f(x,y,V_z)\frac{dV_z}{dz}
\end{equation}
The relation between $z$ and $V_z$ becomes particularly simple in the
case of pure radial expansion, the velocity becoming a simple scalar $V$:
\begin{equation}
  z/r=V_z/V
\end{equation}

As $|z|$ needs not to exceed $r$, $V$ must be larger than $|V_z|$ for $z$ to
be obtained from Relation (4). If not, part of the observed Doppler velocity
spectrum is not accounted for by the calculated effective emissivity. But as
long as this is the case, any scalar velocity field will produce an effective
emissivity that reproduces perfectly the observations. Without further
preconception about the nature of the space distribution of the effective
emissivity, nothing else can be said. In the case of evolved stars, a natural
hypothesis that might help with favouring specific solutions of the
de-projection problem is the assumption of axi-symmetry (including spherical
symmetry) of the effective emissivity, which is known to apply to the vast
majority of such stars.

\subsection{Orientation of the star axis}

In order to measure the deviation from axi-symmetry of the effective emissivity
$\rho(x,y,z)$, we define at each space point $(x,y,z)$ the quantity
\begin{equation}
  D(R',z',\omega)=\frac{\rho(R',z',\omega)-<\rho(R',z')>}{\Delta \rho(R',z')}
\end{equation}
where $R'=\sqrt{x'^2+y'^2}$ and where $\Delta \rho(R',z')$ is the rms fluctuation of $\rho(R',z',\omega)$ about its mean over the volume of the data cube contributing to its evaluation (interpolation over 4 pixels and over the relevant range of the Doppler velocities).
In case of perfect axi-symmetry, $D(R',z',\omega)$ cancels
and we define $\chi^2_{axi} = <D^2(R',z',\omega)>$ as a figure of merit measuring the
deviation from axi-symmetry. $\Delta \rho(R',z')$ accounts well for the dependence over the meridian plane of both experimental and systematic uncertainties. However, it overestimates them by a factor $\sim$3 and $\chi^2_{axi}$ is accordingly smaller than unity by a factor $\sim$10.

\begin{figure}
  \centering
  \includegraphics[width=0.65\textwidth,trim=0.cm 2.2cm 1.cm 1.9cm,clip]{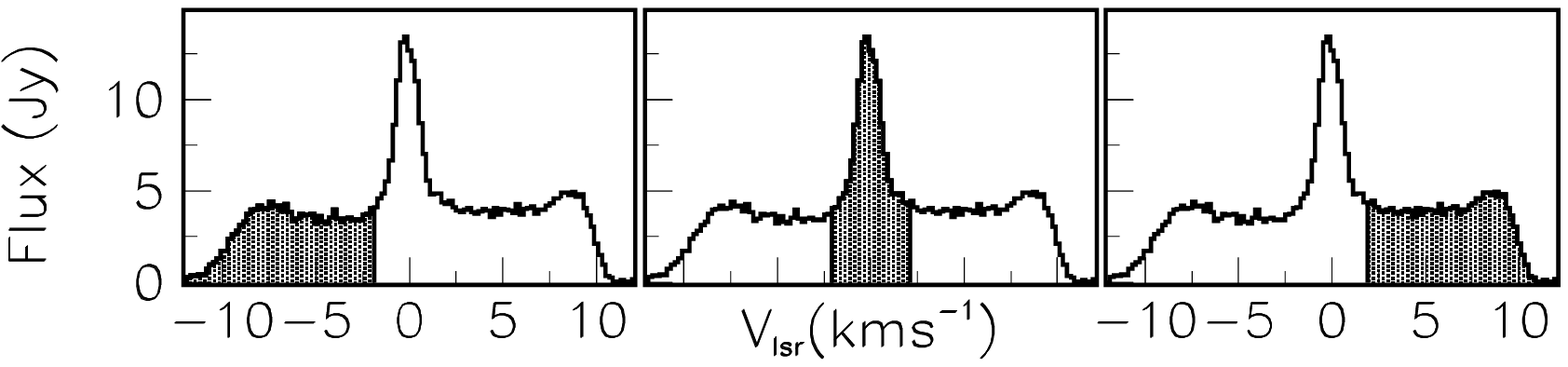}
  \includegraphics[width=0.65\textwidth,trim=0.cm 2.2cm 1.cm 1.7cm,clip]{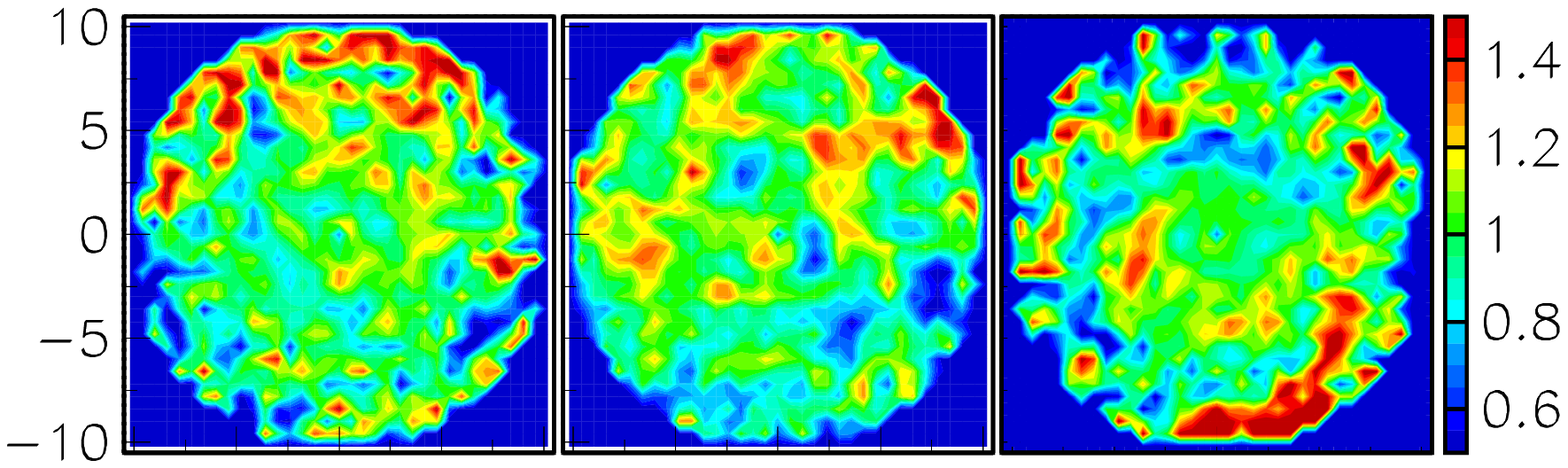}
  \includegraphics[width=0.65\textwidth,trim=0.cm 2.2cm 1.cm 1.7cm,clip]{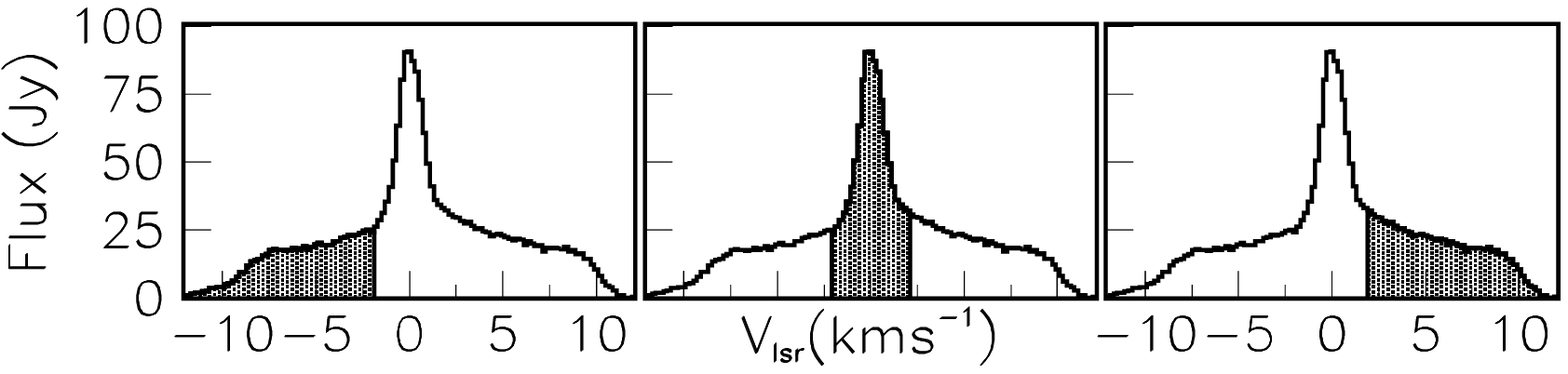}
  \includegraphics[width=0.65\textwidth,trim=0.cm 1.5cm 1.cm 1.7cm,clip]{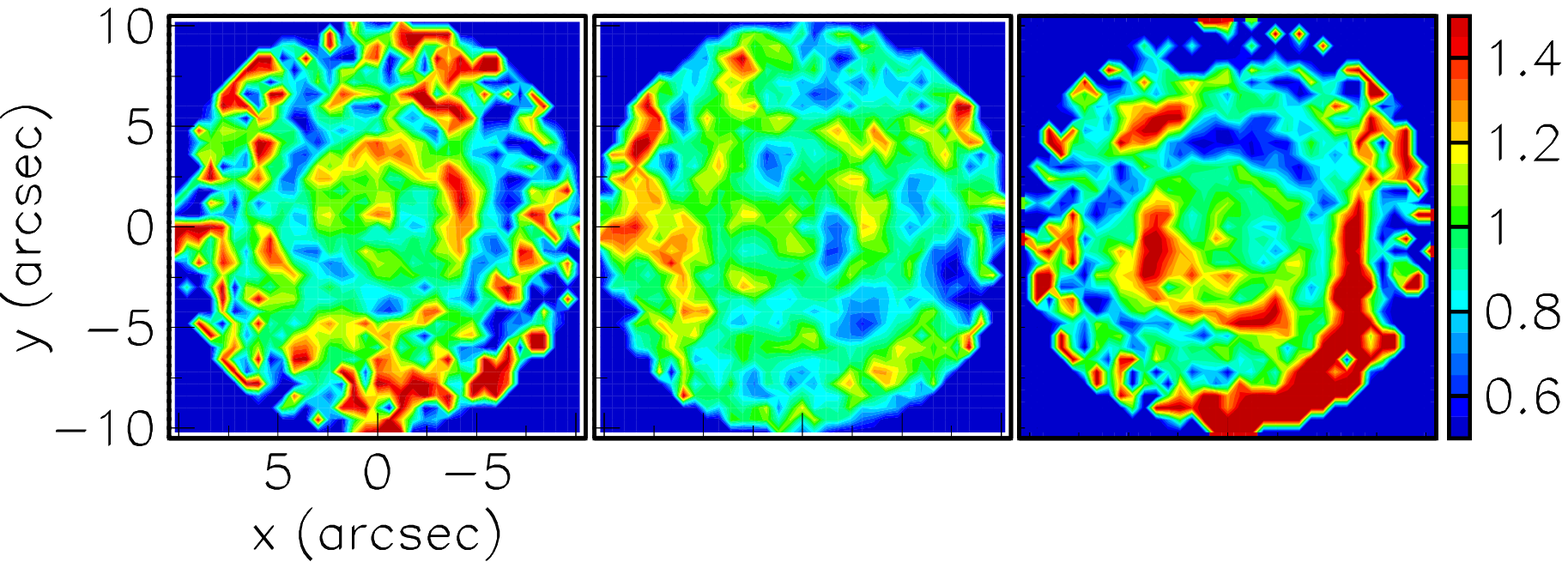}
  \caption{Sky maps of $\Delta(R,\psi)$ for $R<$10 arcsec for \mbox{CO(1-0)} (up) and \mbox{CO(2-1)} (down) emissions.
    Left panels are for $V_z<-2$ \kms, central panels for $|V_z|<$2 \kms\ and right panels for $V_z>$2 \kms\, 
    as indicated in the Doppler velocity spectra displayed above the sky maps (This figure is reproduced from Figure 13,
    \textit{De-projection of radio observations of axisymmetric expanding circumstellar envelopes}, Nhung et al., 2018, MNRAS, 408, 3324).}
  \label{fig4}
\end{figure}

We have calculated the values taken by $\chi^2_{axi}$ for a very broad range of wind
  velocity configurations, from bipolar to equatorial, with or without radial velocity
  gradients, and we find that as long as $\varphi$ is kept small the de-projected effective
  emissivity is axi-symmetric. The reason is an intrinsic axi-symmetry of the data
  about an axis making a small angle with the line of sight for which we give evidence
  below. The channel maps of the measured brightness displayed in Figures \ref{fig2} and \ref{fig3},
  approximately rotation-invariant about the origin, already indicate the existence
  of such symmetry. But the study of the dependence of $\chi^2_{axi}$ on $\varphi$ and $\theta$ tells us in
  addition that the intrinsic symmetry is not spherical. In case of spherical symmetry,
  $\varphi$, and a fortiori $\theta$, are undefined, which is not the case for non-spherical
  axi-symmetry. To illustrate this important result in some detail, we show in
  Figure \ref{fig4} a simplified version of Figures \ref{fig2} and \ref{fig3} aimed at illustrating the
  rotational invariance about the origin; the whole Doppler velocity range is
  split in only three intervals, the blue-shifted wing, the narrow central
  component and the red-shifted wing; moreover, rather than plotting the intensity,
  we plot the ratio \mbox{$\Delta(R,\psi)=F(R,\psi)/<F(R,\psi)>_R$} where $<F(R,\psi)>_R$ is the mean value of
  $F(R,\psi)$ averaged over $\psi$ at fixed $R$. The rms values of $\Delta(R,\psi)-1$, which measure the
  violation of rotation invariance, are listed in Table \ref{table2};
they are dominated by large distances from the star
and never exceed 21\% for \mbox{CO(1-0)} and 35\% for \mbox{CO(2-1)} if one requires $R$ not
to exceed 8 arcsec. There are obvious similarities between the maps of both
emissions, giving confidence in the reality of the features that they may reveal.

\begin{table}
\centering
\caption{Rms deviation from unity (\%) of $\Delta(R,\psi)$}
\begin{tabular}{|c|c|c|c|c|}
\hline
\multirow{2}{*}{} &\multicolumn{2}{c|}{$R<10$ arcsec} & \multicolumn{2}{c|}{$R<8$ arcsec}             \\
\cline{2-5}             
                                      &CO(1-0)  &CO(2-1)  &CO(1-0)  &CO(2-1)   \\
\hline              
                  $V_z<-2$ \kms       &30       &56       &20       &35        \\
\hline              
                  $|V_z|<2$ \kms      &20       &26       &18       &21        \\
\hline              
                  $V_z>2$ \kms        &32       &70       &21       &35        \\
\hline
\end{tabular}\\
\label{table2}
\end{table}

\begin{figure}
\centering
\includegraphics[width=0.7\textwidth,trim=1.5cm 8.5cm 0.cm 2.cm,clip]{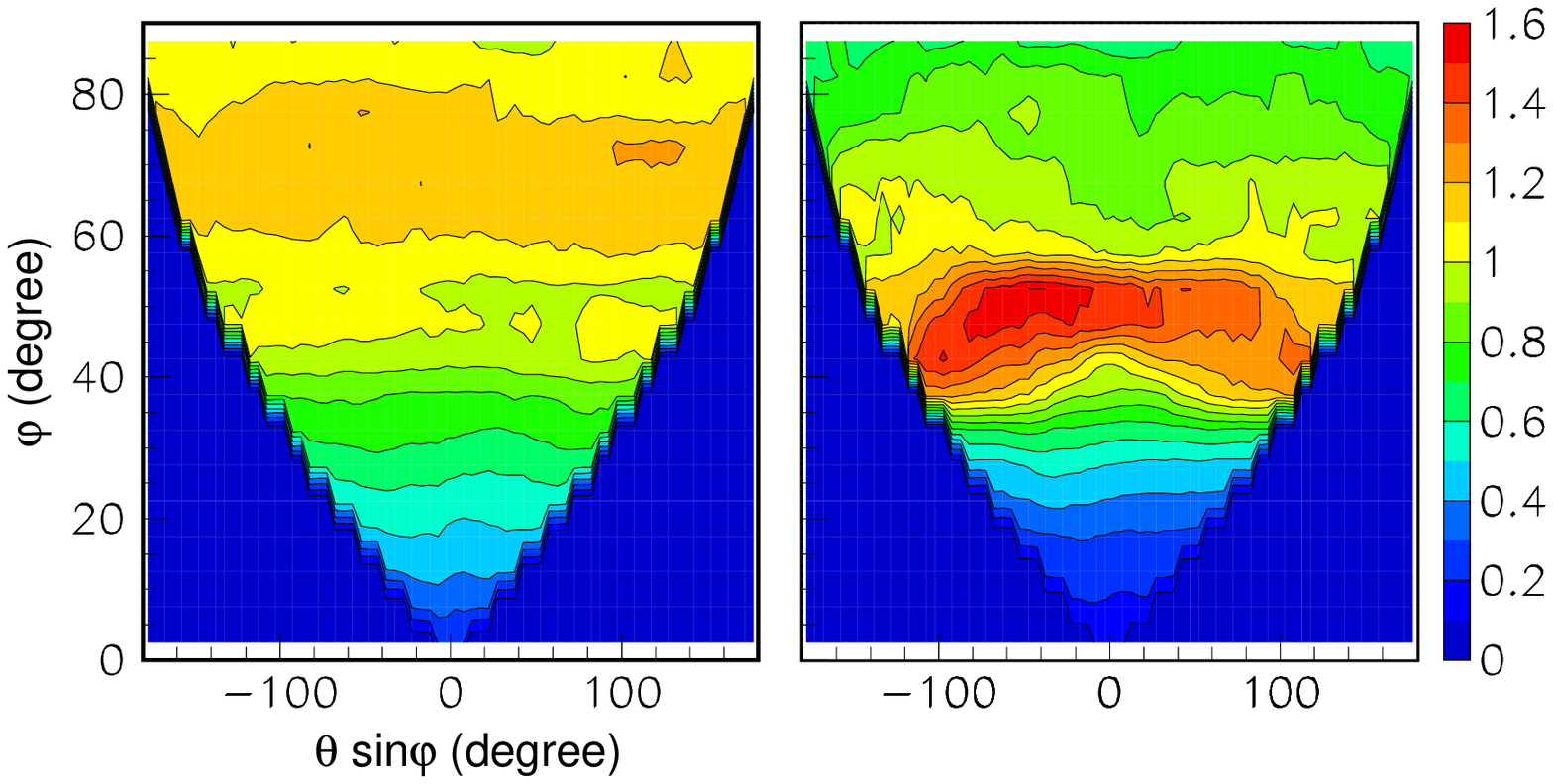}
\includegraphics[width=0.7\textwidth,trim=1.5cm 8.5cm 0.cm 2.cm,clip]{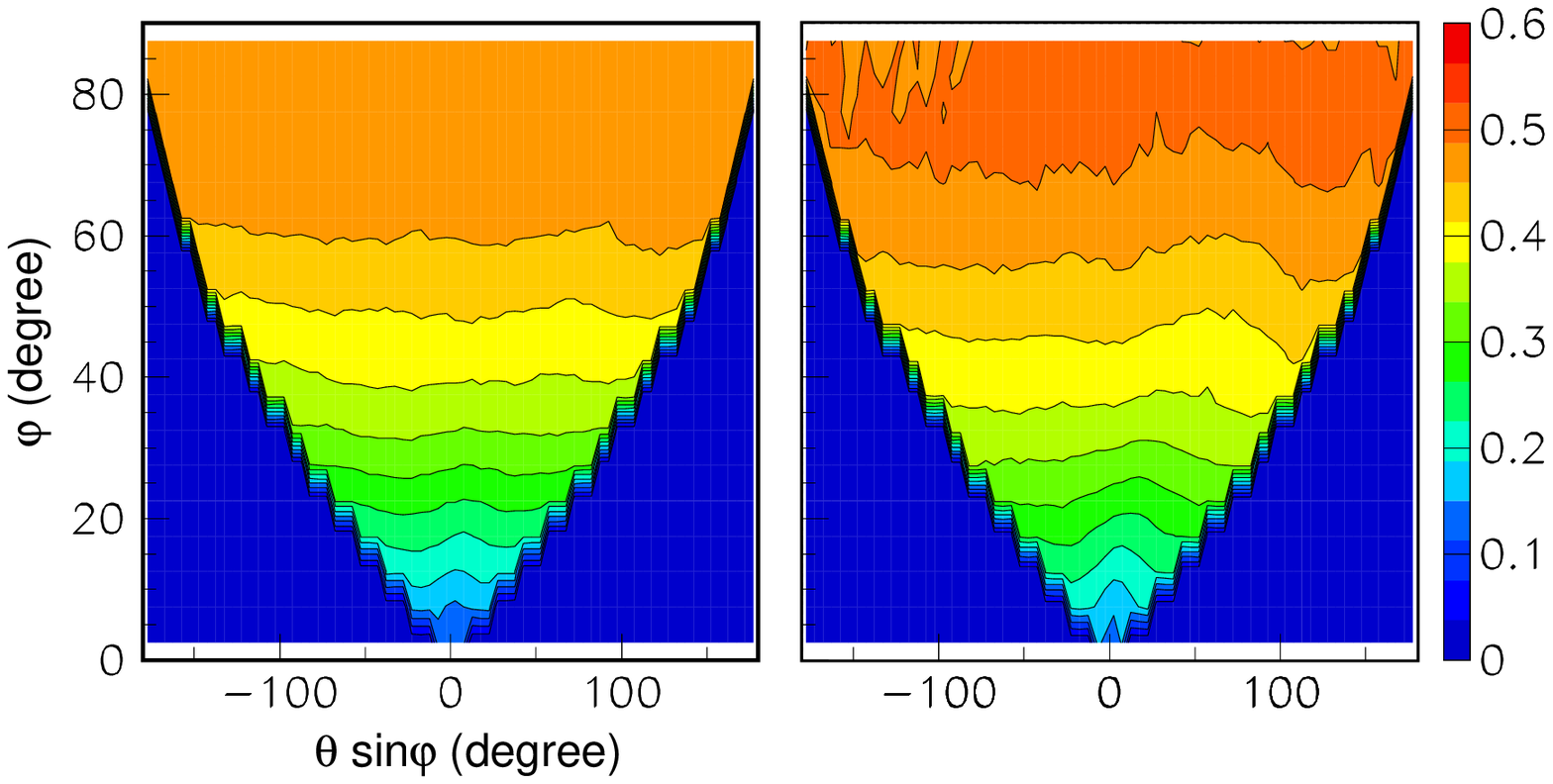}
\caption{Dependence of $\chi^2_{axi}$ on $\theta \sin\varphi$ (abscissa)
  and $\varphi$ (ordinate) for a spherical wind $V_0$=12.5 \kms\ (upper panels) or increasing from 0 to 12.5 \kms\
    between $r=0$ and $r=9$ arcsec (lower panels) and for \mbox{CO(1-0)}
    (left) and \mbox{CO(2-1)} (right) emissions. We use $\theta \sin\varphi$ rather than
    $\theta$ to have bins of constant solid angle coverage, $d\Omega=\sin\varphi d\theta d\varphi$.}
\label{fig5}
\end{figure}

  As was remarked above, Figure \ref{fig4} gives evidence for axi-symmetry about an axis close to
  the line of sight, but does not exclude spherical symmetry; on the contrary, it might
  seem more natural to see it as evidence for spherical symmetry: non-spherical axi-symmetry
  requires a symmetry axis nearly aligned with the line of sight, which one might feel is
  unlikely, even if such a feeling has no meaning (any pre-determined orientation is unlikely).
  However, the study of the dependence of $\chi^2_{axi}$ on $\varphi$ and $\theta$ invalidates such an interpretation.
  What $\chi^2_{axi}$ does is to measure the axi-symmetry of the de-projected effective
  emissivity about the axis of the axi-symmetric wind velocity field used for de-projection : the smaller $\chi^2_{axi}$, the better the axi-symmetry.
  As was mentioned above, in all cases that we have studied, we find that $\chi^2_{axi}$
  is minimal at or near $\varphi=0$. In case of spherical symmetry, one would find
  instead that $\chi^2_{axi}$ is independent of $\varphi$.

  To illustrate this point, we choose a spherical wind configuration to de-project the
  effective emissivity. In such a case, any axi-symmetry of the de-projected emissivity
  is an intrinsic property of the observations. When choosing an orientation ($\theta,\varphi$)
  of the wind axis used for de-projection, it is convenient to use bins of constant
  solid angle, $\varphi$ vs $\theta \sin \varphi$, accounting for the fact that when $\varphi=0$,
  $\theta$ is undefined. Figure \ref{fig5} displays the result and gives evidence for a relatively
  steep minimum near $\varphi=0$ for both CO(1-0) and CO(2-1) emissions. The value of $\varphi$
  for which $\chi^2_{axi}$ reaches twice its minimal value is 5$^\circ$ for both emissions,
  giving the scale of the acceptable range of $\varphi$ values.

  One might wonder if this conclusion may be influenced by the fact that we use wind
  configurations having velocities independent from the distance $r$ to the star. Figure \ref{fig5}
  shows in the lower panels that such is not the case. The wind configuration used for
  de-projection is spherical but assumes a constant positive radial velocity gradient,
  the wind velocity increasing from 0 at $r=0$ to 12.5 \kms\ at $r=9$ arcsec and staying
  there for larger values of $r$. The values of $\varphi$ for which $\chi^2_{axi}$ reaches
  twice its minimal value are now respectively 13$^\circ$ and 15$^\circ$ for CO(1-0) and
  CO(2-1) emissions. Indeed there exist arguments in favour of positive velocity gradients:
  the analysis of Nhung et al. (2015a) gives evidence for these and one might argue that
  the importance over the sky plane of the low velocity narrow central component of the
  Doppler velocity spectrum, rather than being due to an enhancement of the effective
  emissivity in the equatorial region, might rather be due to the presence of a spherical
  volume of gas in very slow expansion (this issue is addressed in some detail later in the text).

  In the remainder of the article, we shall take $\varphi=0$, meaning that $\theta$ is
  irrelevant and that $(x',y',z')=(x,y,z)$, $R'=R$ and $\omega=\psi$. This choice is
  justified to the extent that the arguments that will be developed are not affected by
  small deviations, say below 20$^\circ$, of the value taken by $\varphi$. It is not
  always the case; as we shall see in Section 3.5, polar latitudes prefer $\varphi=0$
  but equatorial latitudes prefer $\varphi\sim 4^\circ$; more generally, independently
  from the adopted wind models, the narrow central Doppler velocity component shows
  evidence for axi-symmetry about an axis making an angle $\varphi$ with the line of
  sight of up to some 10$^\circ$ to 15$^\circ$ and projecting on the sky plane at a position
  angle of some 20$^\circ$ to 30$^\circ$ west of north.
     
\subsection{Bipolar, spherical and equatorial winds}

\begin{figure}
\centering
\includegraphics[width=0.6\textwidth,trim=0.cm 1.cm 1.cm 1.cm,clip]{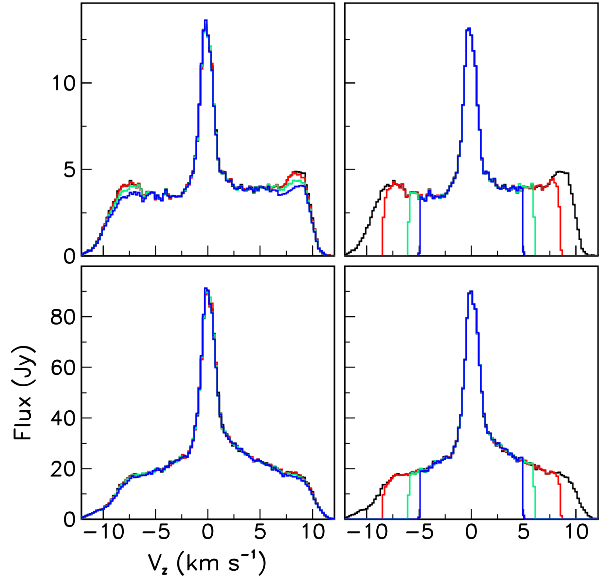}
\caption{ Doppler velocity spectra obtained from the de-projected effective emissivity for \mbox{CO(1-0)} (up)
  and \mbox{CO(2-1)} (down) emissions and bipolar (left) or equatorial (right) winds. In all cases the larger of
  $V_{pole}$ and $V_{eq}$ is set at 12.5 \kms\ and the other takes values of 12.5 (black), 8.5 (red), 4.5 (green)
  and 0.5 (blue) \kms\ respectively. The black histogram, corresponding to a spherical wind configuration,
  is undistinguishable from the measurement.}
\label{fig6}
\end{figure}

The following discussions use models of the wind velocity assuming radial expansion,
independent of the distance $r$ to the star, of the form
\begin{equation}
  V=V_{eq}+(V_{pole}-V_{eq})\sin^2\alpha
\end{equation}
where $V_{pole}$ is the polar and $V_{eq}$ the equatorial velocity. For $V_{eq}=0$
we obtain a bipolar wind $V=V_{pole} \sin^2\alpha$, for $V_{pole}=V_{eq}$ a
spherical wind and for $V_{pole}=0$ an equatorial wind $V=V_{eq} \cos^2\alpha$.
In general $V_{pole}>V_{eq}$ produces bipolar winds, $V_{eq}>V_{pole}$ produces
equatorial winds. In the spherical wind case, writing $V_0=V_{pole}=V_{eq}$
and $\lambda = V_z/V_0$, an effective emissivity of the form
$\rho = \rho_0 r^{-2}$ produces a Doppler velocity spectrum of the form
$f_{sph} = \rho_0 V_0^{-1}R^{-1}(1-\lambda^2)^{-\frac{1}{2}}$ which displays no
narrow central component and ends abruptly at  $|V_z|=V_0$ below which values
it is strongly enhanced.
It takes an enhancement of the effective emissivity in the equatorial plane
to produce a central velocity component. When the wind velocity departs from
spherical, these general features are conserved.
However, the difference
between bipolar and equatorial winds is that the large values of $|V_z|$,
being probed preferentially near the poles, require large values of $V_{pole}$,
not of $V_{eq}$. This is illustrated in Figure~\ref{fig6}, which displays the
Doppler velocity spectrum accounted for by the de-projected effective
emissivity for different values of the model parameters. Bipolar and spherical
winds do not require $V_{pole}$ to exceed significantly 12.5 \kms\ while
equatorial winds are severely truncated as soon as $V_{pole}$ decreases below
this value. For the whole Doppler velocity spectrum to be accounted for by
an equatorial wind, we need to keep $V_{pole}$ sufficiently large: for a given
shape of the wind velocity distribution, meaning a given $V_{eq}/V_{pole}$
ratio, we need to increase $V_{eq}$ accordingly. Examples are given in the next
section. In other words, bipolar winds require smaller wind velocities than
equatorial winds do.

\begin{figure*}
\centering
\includegraphics[height=4.75cm,trim=0.5cm 1.5cm .5cm 1.cm,clip]{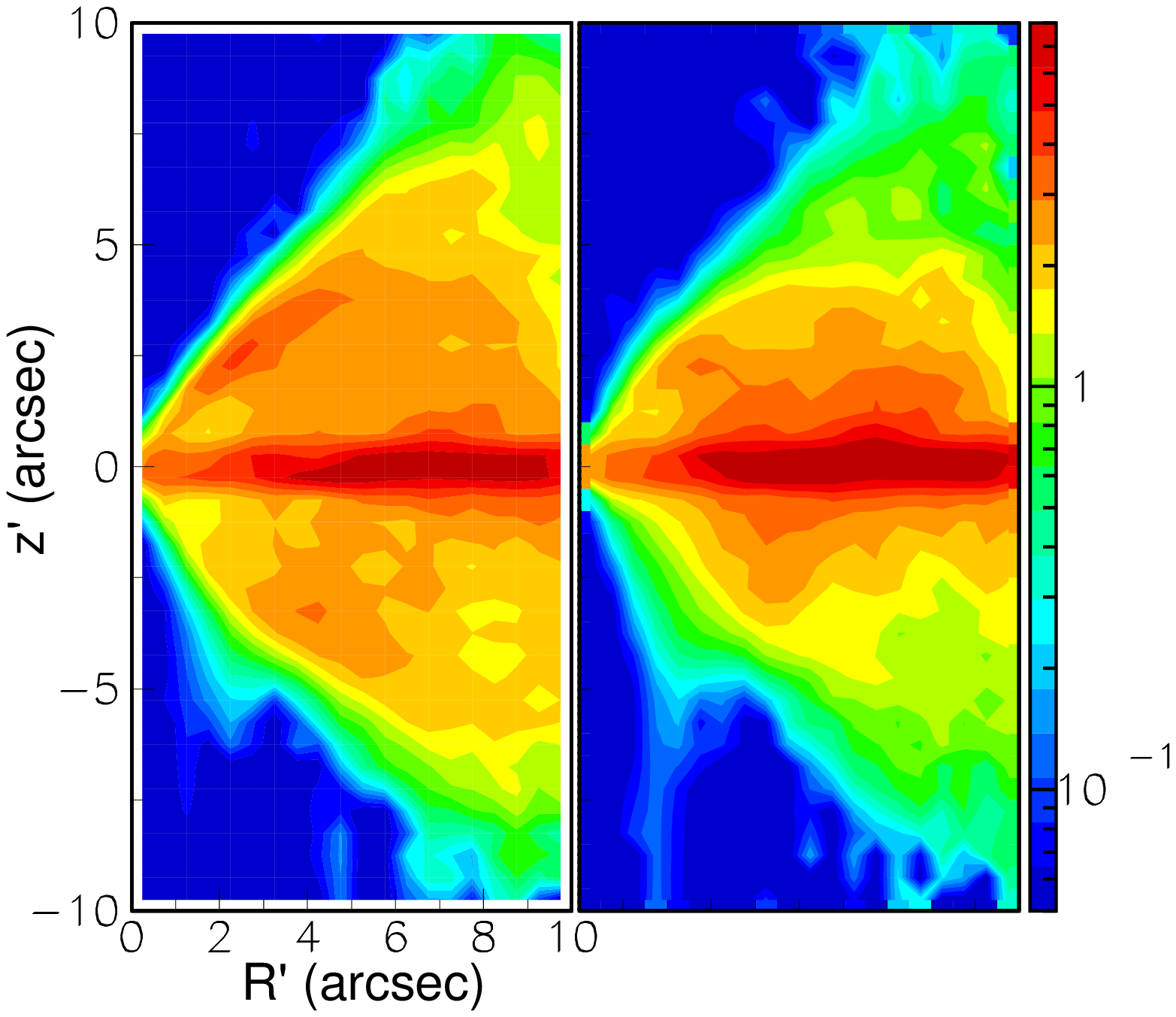}
\includegraphics[height=4.75cm,trim=0.5cm 1.5cm 2.cm 1.cm,clip]{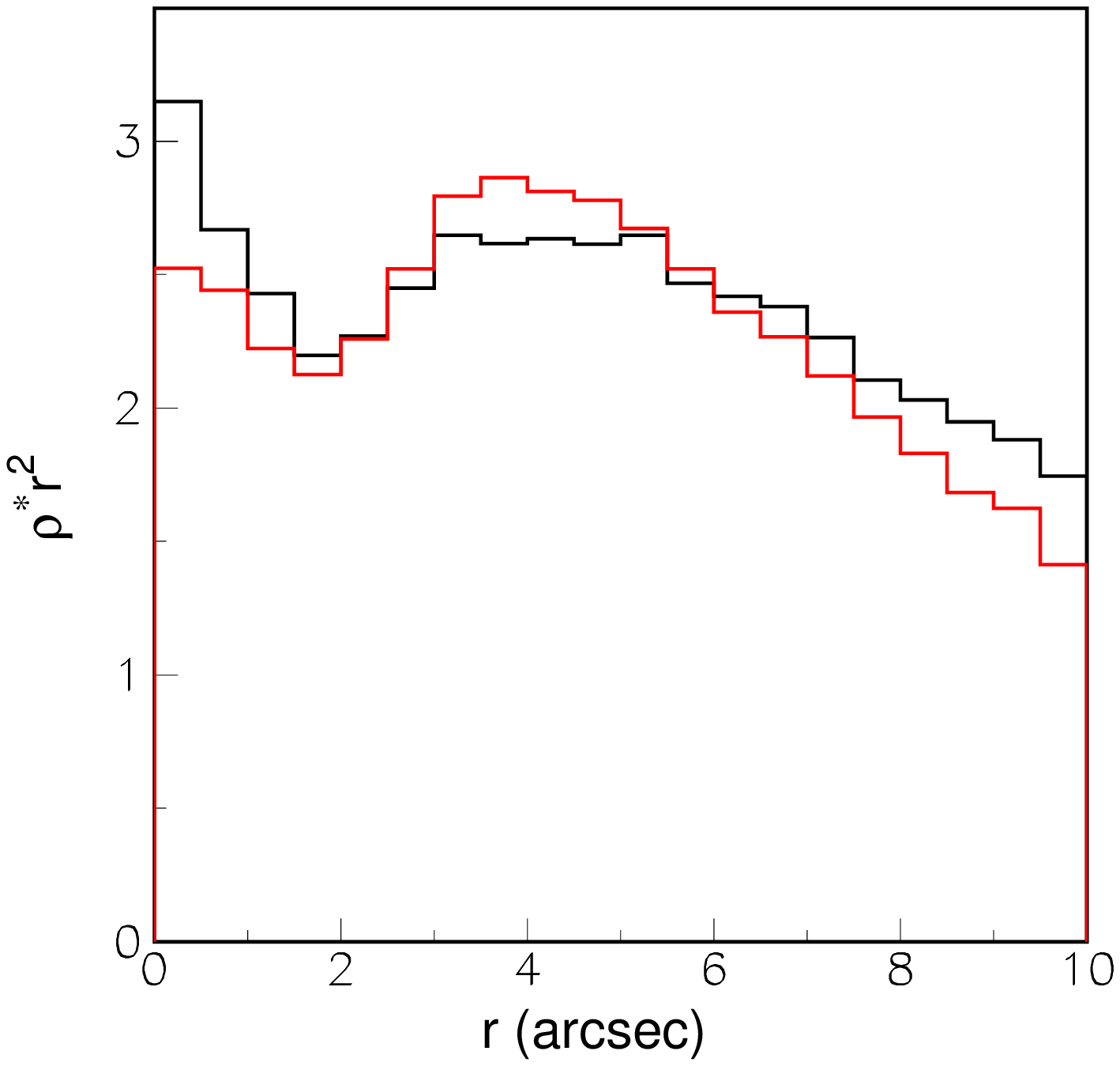}
\includegraphics[height=4.75cm,trim=1.5cm 1.5cm 2.5cm 1.cm,clip]{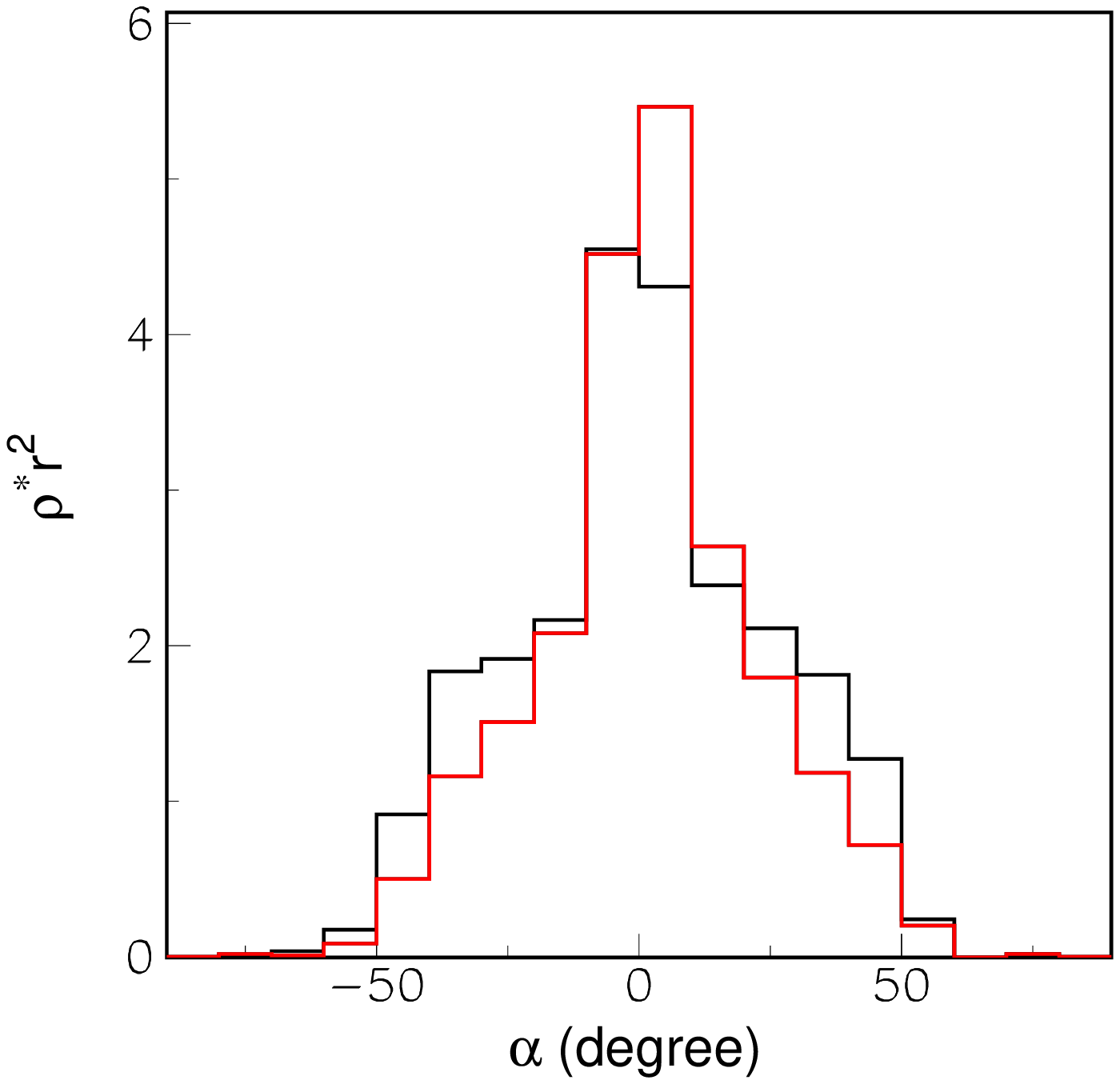}
\caption{ De-projected temperature-corrected effective emissivity $\rho^*$ (see Section 3.4)
  multiplied by $r^2$ obtained from a spherical radial wind of velocity 12.5 \kms. The half-meridian
  plane maps, from left to right \mbox{CO(1-0)} and \mbox{CO(2-1)} emissions, are shown in the left panels.
  The dependence on $r$ (middle) and on $\alpha$ (right) is shown for \mbox{CO(1-0)} (black) and \mbox{CO(2-1)}
  (red) in the right panels. The quantity $\rho^*r^2$ is in units that correspond approximately
  to a flux of matter of $\sim$10$^{-8}$ solar masses per steradian and per year for an expansion
  velocity of 10 \kms\ or, equivalently, to a flux of matter of $\sim$0.5\,10$^{-8}$ solar masses per
  steradian and per AU. A more precise evaluation, accounting for the effect of absorption,
  is deferred to a forthcoming paper (Hoai et al., in preparation).}
\label{fig7}
\end{figure*}

The end points of the measured Doppler velocity spectra, being much less
abruptly cut-off than a $f_{sph}= \rho_0 V_0 ^{-1}R^{-1}(1-\lambda^2)^{-\frac{1}{2}}$
distribution, imply that the effective emissivity needs to be depressed near
the poles. Moreover, the presence in the Doppler velocity spectra, everywhere
on the sky plane, of a strong central narrow component requires an enhancement
of the effective emissivity near the equator. This is clearly visible in
Figure~\ref{fig7}, which displays the de-projected effective emissivity
obtained from a spherical wind having a radial velocity $V_0$=12.5 \kms.

The main features of the de-projected effective emissivity observed for
a spherical wind, strong polar depression and equatorial enhancement, remain
qualitatively present for other wind models for the same reasons as have been
invoked in the spherical case. Also, the presence of a maximum at distances of
some 3 to 6 arcsec from the star, will be seen in the next section to persist
to some level over a broad range of models. However, introducing positive
  radial velocity gradients increases the contribution of a large and very slowly
  expanding gas volume to the narrow central component of the Doppler velocity
  spectrum, thereby lessening the need for an equatorial enhancement of the effective emissivity.

\subsection{Dependence of the flux of matter on $r$ and $\alpha$}

In the present section, we calculate the de-projected effective emissivity
for different representative models of the wind velocity. From what was learnt
in the preceding section, we always adjust the larger of $V_{pole}$ and $V_{eq}$
in such a way as to account for the whole observed Doppler velocity spectrum,
but not larger than necessary. While sensible, this procedure cannot be
rigorously justified. Moreover, we introduce a new parameter $n$ that allows
for changing the angular width of the bipolar or equatorial outflow by writing
\begin{align*}
  V&=V_{eq}+(V_{pole}-V_{eq})(\sin^2\alpha)^n \, \mbox{for $V_{pole}>V_{eq}$ (bipolar)}\\
  V&=V_{pole}+(V_{eq}-V_{pole})(\cos^2\alpha)^n \, \mbox{for $V_{eq}>V_{pole}$ (equatorial)}
\end{align*}

For $n$=1, as used in the preceding section, the half-width at half-maximum
of the enhancement is 45$^\circ$. It is 60$^\circ$ for $n$=0.5 and 30$^\circ$
for $n$=2.4. Our choice of parameters,
  meant to cover a very broad range of wind velocity configurations from strongly
  prolate (bipolar) to strongly oblate (equatorial),  is listed in Table~\ref{table3}
together with the value of $V_{pole}$ or $V_{eq}$ adjusted to just account for
the whole observed Doppler velocity spectra.

\begin{table}
\centering
\caption{ Parameters of the representative wind models.}
\begin{adjustbox}{max width=\textwidth}
\begin{tabular}{|c|c|c|c|c|c|c|c|}
\hline
\multicolumn{2}{|c|}{$n$} &\multicolumn{2}{c|}{0.5}   &\multicolumn{2}{c|}{1}   &\multicolumn{2}{c|}{2.4}       \\
\hline
\multirow{2}{*}{Polar}   &$V_{pole}$       &12   &12   &13.2  &12   &16.8 &12.4 \\
\cline{2-8}
&$V_{eq}/V_{pole}$&1/4  &3/4  &1/4   &3/4  &1/4  &3/4  \\
\cline{2-8}
&$V_{eq}$         &3    &9    &3.3   &9    &4.2  &9.3  \\
\hline
\multirow{2}{*}{Equatorial}&$V_{eq}$ &28.5 &13.6 &4.5   &14.4 &36   &14.8 \\
\cline{2-8}
                                    &$V_{pole}/V_{eq}$&1/3  &3/4  &1/3   &3/4  &1/3  &3/4  \\
\cline{2-8}
                           &$V_{pole}$       &9.5  &10.2 &11.5 &10.8 &12   &11.1 \\
\hline
\end{tabular}
\end{adjustbox}
\label{table3}
\end{table}

\begin{figure*}
\centering
\includegraphics[height=10.cm,trim=2.cm 1.5cm 2cm 1.cm,clip]{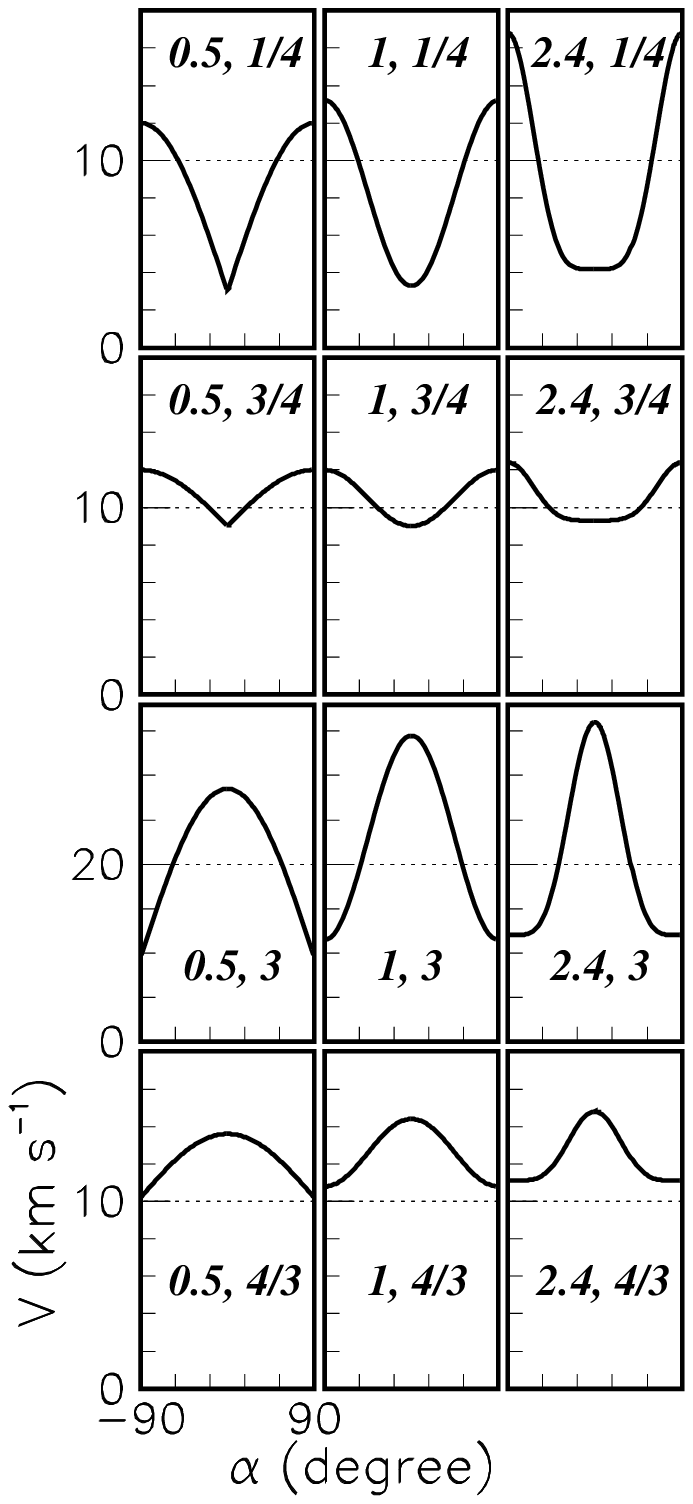}
\includegraphics[height=10.cm,trim=2.1cm 1.5cm 2cm 1.cm,clip]{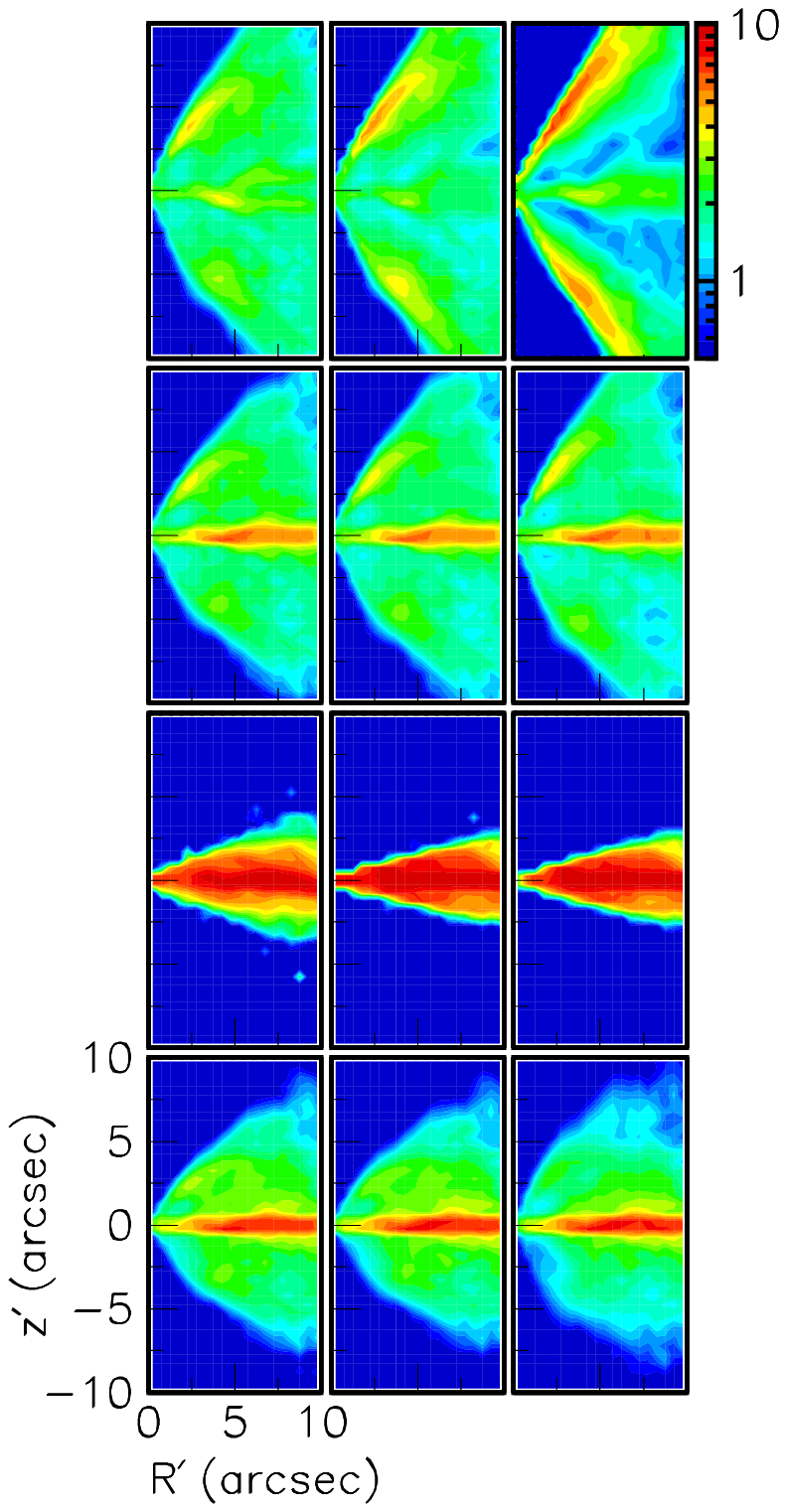}
\includegraphics[height=10.cm,trim=2.1cm 1.5cm 1.9cm 1.cm,clip]{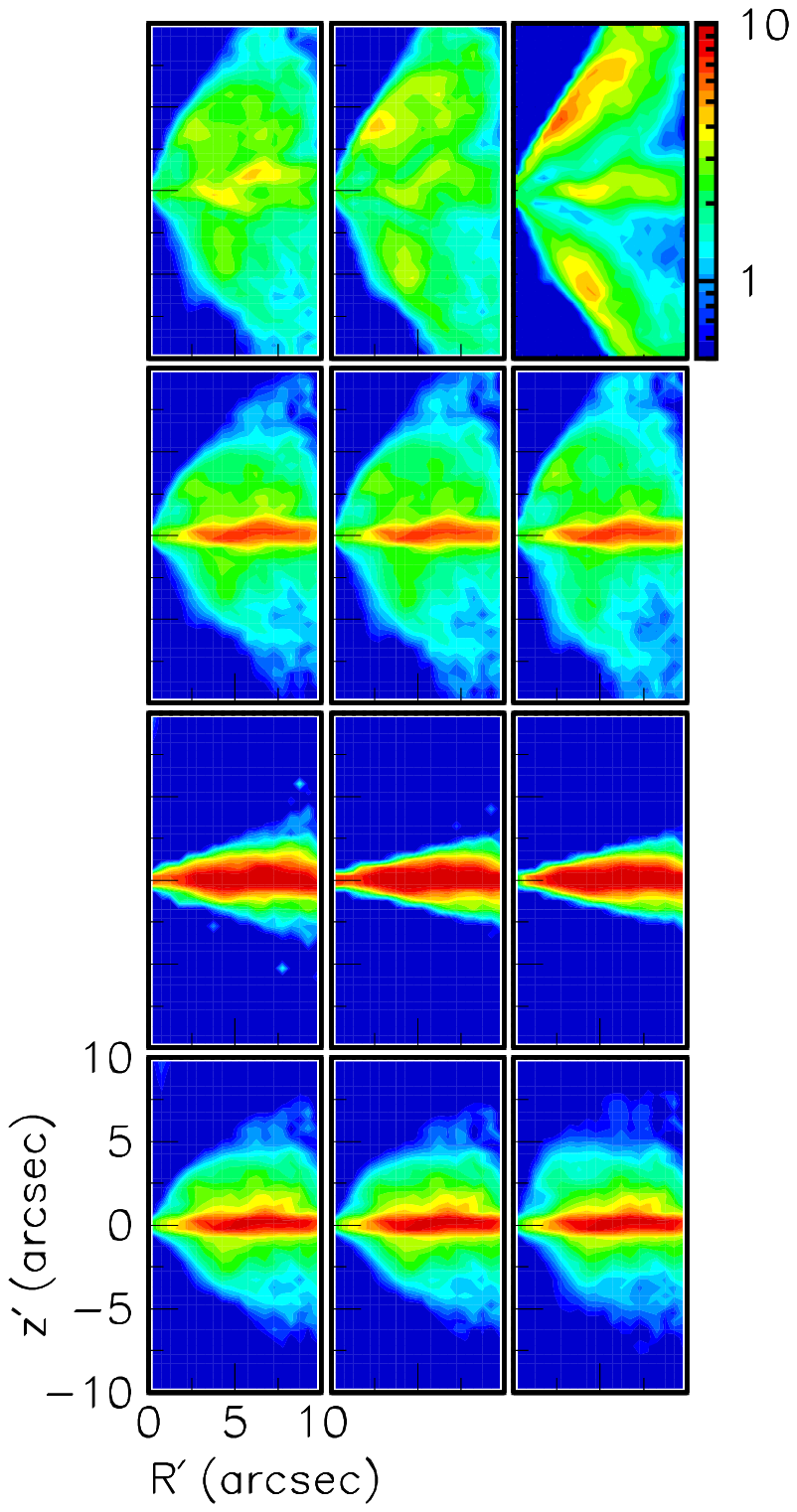}
\caption{ Left: dependence on latitude $\alpha$ (degrees in abscissa) of the wind radial velocity $V$ (\kms\ in ordinate) for each of the twelve combinations $(V_{pole},V_{eq},n)$ of the model parameters as indicated in each sub-panel in the form of $n,V_{eq}/V_{pole}$. Right: Half-meridian-plane maps, averaged over longitude, of the de-projected effective emissivity corrected for temperature, $\rho^*$, multiplied by $r^2$ for each of the twelve combinations of the model parameters, ordered as in the left panel, and for \mbox{CO(1-0)} and \mbox{CO(2-1)} emissions separately (left to right). Units are as described in the caption of Figure \ref{fig7}.}
\label{fig8}
\end{figure*}

\begin{figure*}
\centering
\includegraphics[width=0.49\textwidth,trim=1.cm 1.5cm 2.cm 1.cm,clip]{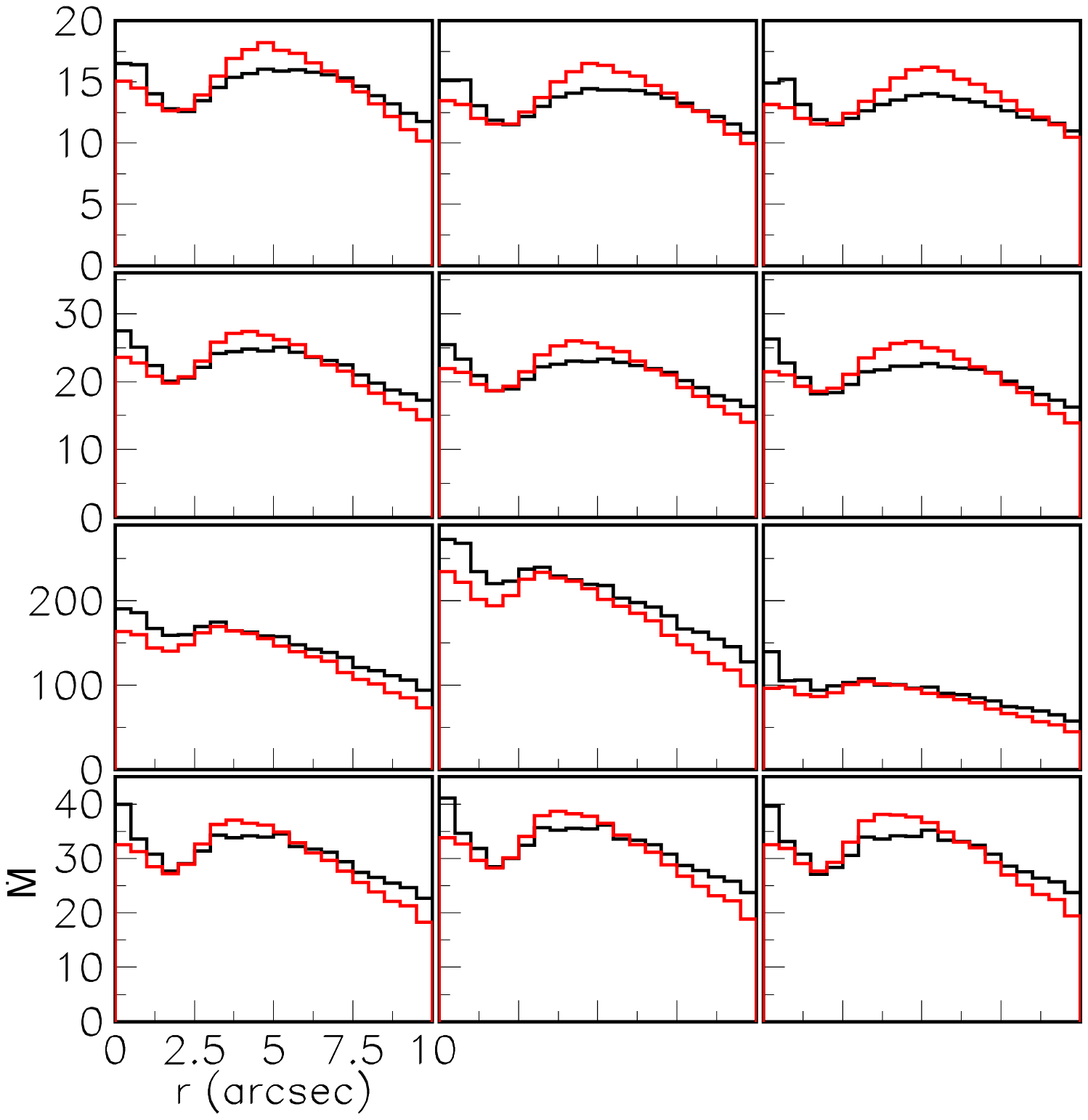}
\includegraphics[width=0.49\textwidth,trim=1.cm 1.5cm 2.cm 1.cm,clip]{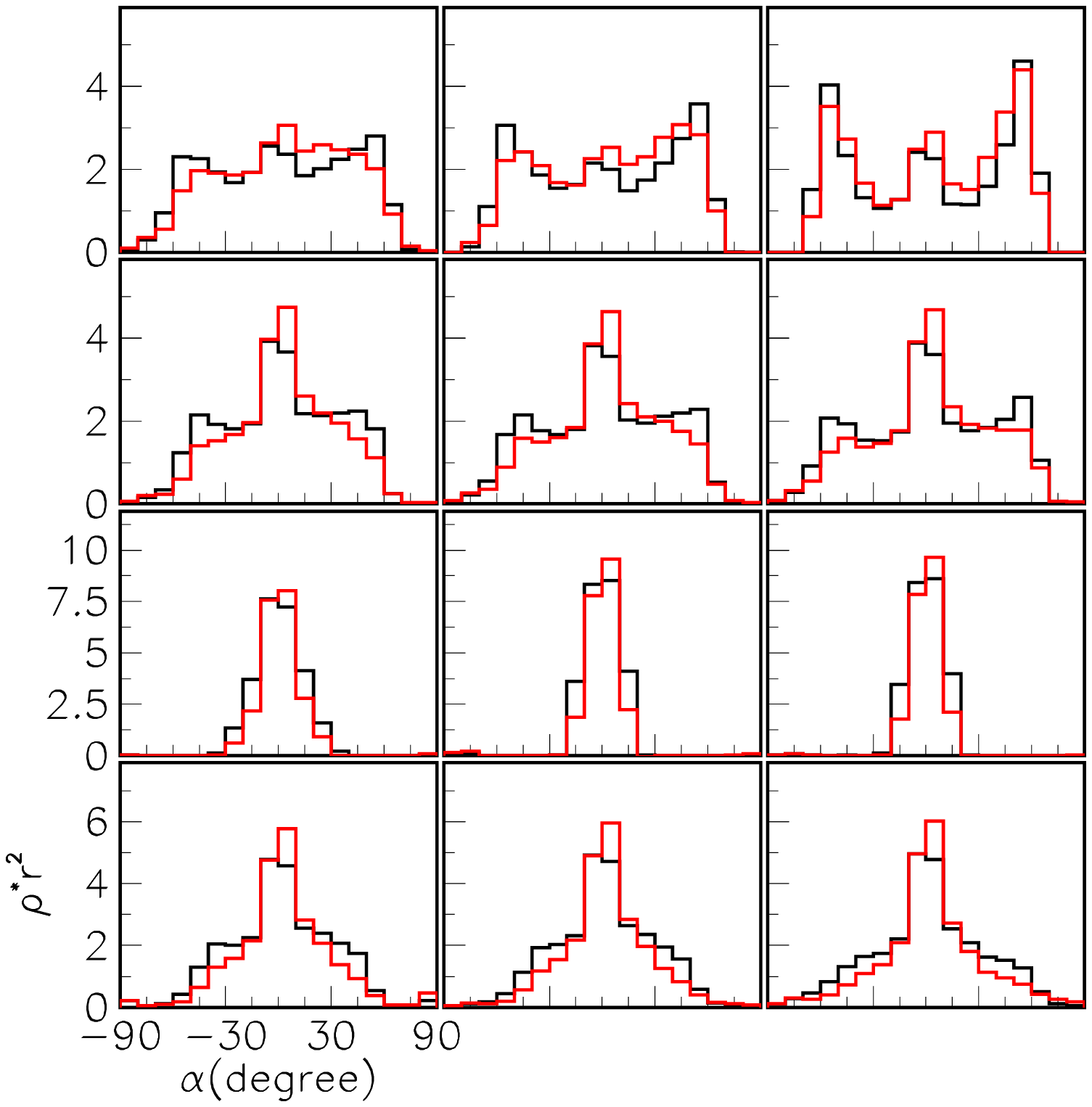}
\caption{ Dependence of the flux of matter on distance $r$ to the star (arcsec, left) and of
  the product $\rho^*r^2$ on latitude $\alpha$ (degrees, right) for each of the twelve combinations
  of model parameters, ordered as in Figure \ref{fig8}. Black is for \mbox{CO(1-0)} emission and red for \mbox{CO(2-1)}
  emission. Units are as described in the caption of Figure \ref{fig7}.}
\label{fig9}
\end{figure*}

Figure~\ref{fig8} (left) displays the corresponding dependence on latitude of
the wind radial velocity $V$. In all cases (Figure~\ref{fig8} right), one obtains a very good agreement
between \mbox{CO(1-0)} and \mbox{CO(2-1)} results after correction for temperature. More
precisely, we define temperature-corrected emissivities $\rho^*=\rho/Q$ where, assuming Local Thermal Equilibrium (LTE),
$Q=Q_0(2J+1)(2.8/T)\exp(-E_{up}/T)$; for \mbox{CO(1-0)} or respectively \mbox{CO(2-1)},
$J$=1 or 2, the energy of the upper level of the transition $E_{up}$=5.5 K or 16.6 K and Einstein's spontaneous emission probability $Q_0$=7.4 10$^{-8}$ s$^{-1}$ or
7.1 10$^{-7}$ s$^{-1}$. As radial dependence of the temperature we assume a
form $T=32/r^{0.65}$ K with $r$ measured in arcsec (one of possible forms
obtained in a preliminary radiative transfer calculation). This temperature
correction is irrelevant to what follows, its only merit is to offer a direct
and transparent comparison between the \mbox{CO(1-0)} and \mbox{CO(2-1)} results by making
them commensurate. Figure~\ref{fig9} displays the dependence on $r$ of the
flux of matter $\dot{M}$, approximated by the expression $\dot{M}=\rho^* V r^2$,
and on $\alpha$ of the temperature-corrected effective emissivity $\rho^*$
multiplied by $r^2$. Note that UV dissociation of the CO molecules causes
a decrease of the density (Mamon et al. 1988) of the approximate form
$f_{UV}={\frac{1}{2^q}}$ with $q=(\frac{r}{13})^{2.1}$ ($r$ in arcsec)
obtained by assuming $\dot{M}=10^{-7}$ \(\textup{M}_\odot\)yr$^{-1}$ and $V\sim$7.5 \kms, meaning
a factor $\sim$91\% at $r$=5 arcsec and $\sim\frac{2}{3}$ at $r$=10 arcsec,
close to what is observed.

\subsection{Deviations from axi-symmetry and disc rotation}

\begin{figure}
\centering
\includegraphics[width=0.7\textwidth,trim=1.3cm .5cm .5cm 1.cm,clip]{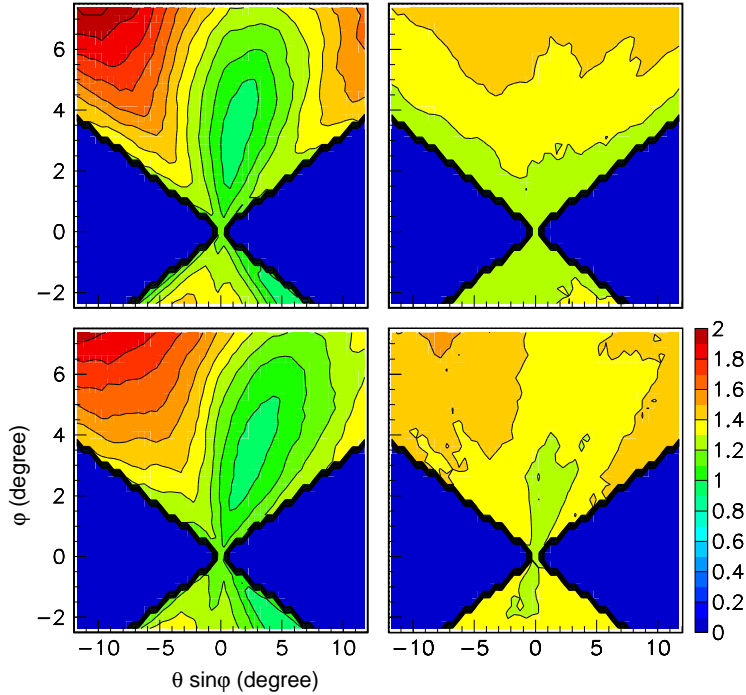}
\caption{ Dependence of $\chi^2_{axi}$ on $\theta \sin\varphi$ (abscissa) and $\varphi$ (ordinate)
  for \mbox{CO(1-0)} (up) and \mbox{CO(2-1)} (down) emissions and for $|z'|<$ 1 arcsec (left) and
  $|z'|>$2 arcsec (right). For convenience, $\chi^2_{axi}$ has been normalised
    to its minimal value and the colour code covers from minimum (1) to twice minimum (2) in each of the four panels.}
\label{fig10}
\end{figure}

The preceding section dealt with data averaged over longitude; we now study
the dependence on longitude of the effective emissivity. For this study we
make a particular choice of parameters meant to be typical of a likely
configuration of velocities and densities:
$(V_{pole},V_{eq},n)$=(12.5, 6, 1.5). Although we do not expect that the 
precise values of $\theta$ and $\varphi$ could matter much, we adjust
them by minimising $\chi^2_{axi}$ in order to be as free as possible of
deviations from axi-symmetry caused by a slight tilt 
of the star axis with respect to the line of sight. We perform the
minimisation separately for the equatorial disc region, $|z'|<$1 arcsec
and the bi-conical outflow region, $|z'|>$2 arcsec. The result is displayed
in Figure~\ref{fig10}. While the latter requires only $\varphi$ to be small,
leaving $\theta$ essentially undefined, the former displays a clear
minimum at $\varphi = 3.8 \pm 2.0^\circ$ where the error corresponds to a 10\% increase in $\chi^2_{axi}$
with excellent agreement between the \mbox{CO(1-0)} and \mbox{CO(2-1)} lines.
The value of $\theta$ at minimum is less well defined and of the order of 35 $\pm$ 35$^\circ$. This is
at strong variance with the behaviour of $\chi^2_{axi}$ in other regions
of space and is commented on below. We therefore adopt these values of
$\theta$ and $\varphi$ to map the effective emissivity, temperature-corrected
and multiplied by $r^2$, in sixteen slabs of $z'$ parallel to the equator,
each 1 arcsec thick and covering between $z'$=$-$8 arcsec and $+$8 arcsec.
The \mbox{CO(1-0)} and \mbox{CO(2-1)} maps display strong similarity; in order to illustrate
it, rather than showing them separately, we prefer to show in Figure~\ref{fig11}
their half-sum and their half-difference. The former gives clear evidence for
the bi-conical morphology of the polar outflow and for the presence of a
flared disc, or of a ring, at the equator. The latter shows that the small
difference between the emissions of the two lines decreases from being
positive at large values of $|z'|$ to negative at the equator, which probably
reveals the imperfection of the radial dependence of the temperature adopted
here. Deviations from axi-symmetry and from symmetry with respect to the
equatorial plane are clearly significant but none of these suggests the
presence of a specific feature worth mentioning; their main characteristic
is their relatively modest amplitude.

The atypical behaviour of $\chi^2_{axi}$ in the disc region may suggest that it
be related to rotation. Rotation of velocity $V_{rot}$ about an axis parallel to
the line of sight implies $V_z$=0 while expansion of velocity $V_{exp}$ implies
$V_z=V_{exp}\, z/r=V_{exp} \sin \alpha$. But, in the equatorial plane, a small tilt
$\varphi$ of the star axis with respect to the line of sight generates Doppler
components $V_z=V_{rot} \cos \omega \sin \varphi$ and
$V_z=V_{exp} \sin \omega \sin \varphi$ : rotation and expansion cancel and
reach their extrema at longitudes differing by 90$^\circ$. Interpreting the
effect of rotation as being caused by expansion produces identical values
of the product $V \sin \varphi$ but position angles $\theta$ at 90$^\circ$ from
each other. However, when moving away from the equatorial plane, the Doppler
component induced by rotation stays constant while that induced by expansion
gets soon dominated by the $V_{exp} \sin \alpha$ term that takes opposite
values in the blue-shifted and red-shifted regions. Moreover, in a plane of
constant $z'$ ($|z'|<<r$) close to the equator, the Doppler component induced by
expansion decreases like $\sim 1/R$ while that induced by rotation would
decrease like $1/R^{\frac{1}{2}}$ in the Keplerian case.

These comments remind us of the difficulty to tell expansion from rotation
(Diep et al. 2016); the study of $\chi^2_{axi}$ is equivalent to that of the
behaviour of the narrow Doppler velocity component over the plane of the sky,
as had been done in Nhung et al. (2015a); it is just more elegant in
summarizing the available information in a single plot. However, a reliable
interpretation in terms of competing expansion and rotation, as is observed,
for example in the Red Rectangle (Tuan-Anh et al. 2015), would probably
require data of better spatial resolution and signal-to-noise ratio and
is beyond the scope of the present article. What has been made clear in the
present analysis is that all the relevant information available on the precise
orientation of the star axis is contained in the behaviour of the narrow
central Doppler velocity component on the plane of the sky, meaning the
kinematics of the equatorial disc or ring. Figure~\ref{fig10} is an elegant
way to summarize its content but the interpretation is made difficult by
the smallness of the tilt angle and a possible competition between rotation
and expansion.

\section{Discussion}
By construction, any combination of model parameters chosen in the region
explored in the preceding section produces an effective emissivity that
gives a perfect description of the observations and that displays approximate
axi-symmetry about the line of sight. Expressing a preference for a particular
model implies therefore arguments of a different nature.

Qualitatively, the dependence on star latitude of the effective emissivity,
and a fortiori of the flux of matter, is found to follow that of the wind
velocity with the exception of the omni-present polar depression that has
been discussed in Section 3.3. Indeed, if $V$ is small, $dV_z/dz$ is also
small and so is $\rho$. As expected, the polar depression is slightly more
important for \mbox{CO(2-1)} than for \mbox{CO(1-0)}, reflecting the difference of behaviour
of the corresponding velocity spectra near their end points. The presence
of the polar depression had been overlooked in previous analyses. Another
important result that had also been essentially overlooked in earlier analyses
is that large equatorial expansion velocities produce a flared disc or ring
of effective emissivity and mass loss. This becomes particularly spectacular
in the case of equatorial winds, for which very high equatorial velocities
may be required. In such cases, the bipolar outflow disappears completely,
the mass loss occurring exclusively near the equator.

Remarkably, the radial dependence of the flux of matter is always
found to display a maximum at distances between 3 and 6 arcsec,
namely $\sim$ 500 AU, from the star. Note that a distance of 10 arcsec,
meaning 1140 AU, is covered in only 540 years at a velocity of 10 \kms:
we see but a very recent past of the history of the dying star.

Such maximum would imply that the assumption of stationary flow, implicit in
the absence of radial dependence of the wind velocity, is not strictly obeyed:
in that case, some episode of increased expansion must have occurred at some
point in the recent history of the star, say in the past 1000 years or so.
Violation of stationarity implies in turn the need to allow for a
  dependence of the wind velocity on the distance $r$ to the star. As mentioned
  earlier, this does not affect the presence of an intrinsic non-spherical
  axi-symmetry (Section 3.2) but lessens the contribution of the equatorial
  effective-emissivity to the central narrow component of the Doppler velocity
  spectrum (Section 3.3). However, the dominance of this narrow component in
  defining the orientation of the star axis (Section 3.5) pleads in favour of
  a slightly tilted equatorial enhancement: a slowly expanding spherical volume
  does not contribute any non-spherical axi-symmetry.

In any case, a significant violation of stationarity is not expected to
  affect the topology of the effective emissivity.
We developed the argument previously
in the case of Mira Ceti (Nhung et al. 2016, Diep et al. 2016) and it is even
stronger in the present case where the morphology of the emission is
considerably simpler, giving strong confidence in the reliability of the
conclusions. Equatorial winds tend to give a steeper decrease of the flux
of matter with distance than bipolar winds do, providing an additional
argument in favour of bipolar winds. However, a detailed analysis in terms
of temperature, optical depth and absorption is required before making
this point more strongly.

In the same vein, we note small occasional differences between the
radial dependence of the \mbox{CO(1-0)} and \mbox{CO(2-1)} flux of matter: these are
unimportant and easily accounted for by small changes in the form adopted
for the radial dependence of the temperature: they do not need to be further
discussed in the present article.

Axi-symmetrical structures can be generated by different mechanisms (e.g., rotation, magnetic fields, binarity, ...).
  For binary systems, Mastrodemos \& Morris (1999) and Theuns \& Jorissen (1993) predict the
  formation of spirals and arc-like structures. A visual inspection of the channel maps in
  Figures \ref{fig2} and \ref{fig3} reveals such sub-structures
  which therefore might hint to the presence of a binary system at the heart of EP Aqr.

The presence of an equatorial ring or flared disc raises the question of its possible rotation.
This issue was addressed at the end of the preceding section and the very small
tilt angle between the intrinsic axis of the observed brightness distribution and
the line of sight precludes a reliable answer to this question.

\begin{figure*}
\centering
\begin{minipage}[c][][t]{1.\textwidth}
\centering
\includegraphics[width=0.49\textwidth,trim=1.5cm 1.5cm 1.5cm 1.cm,clip]{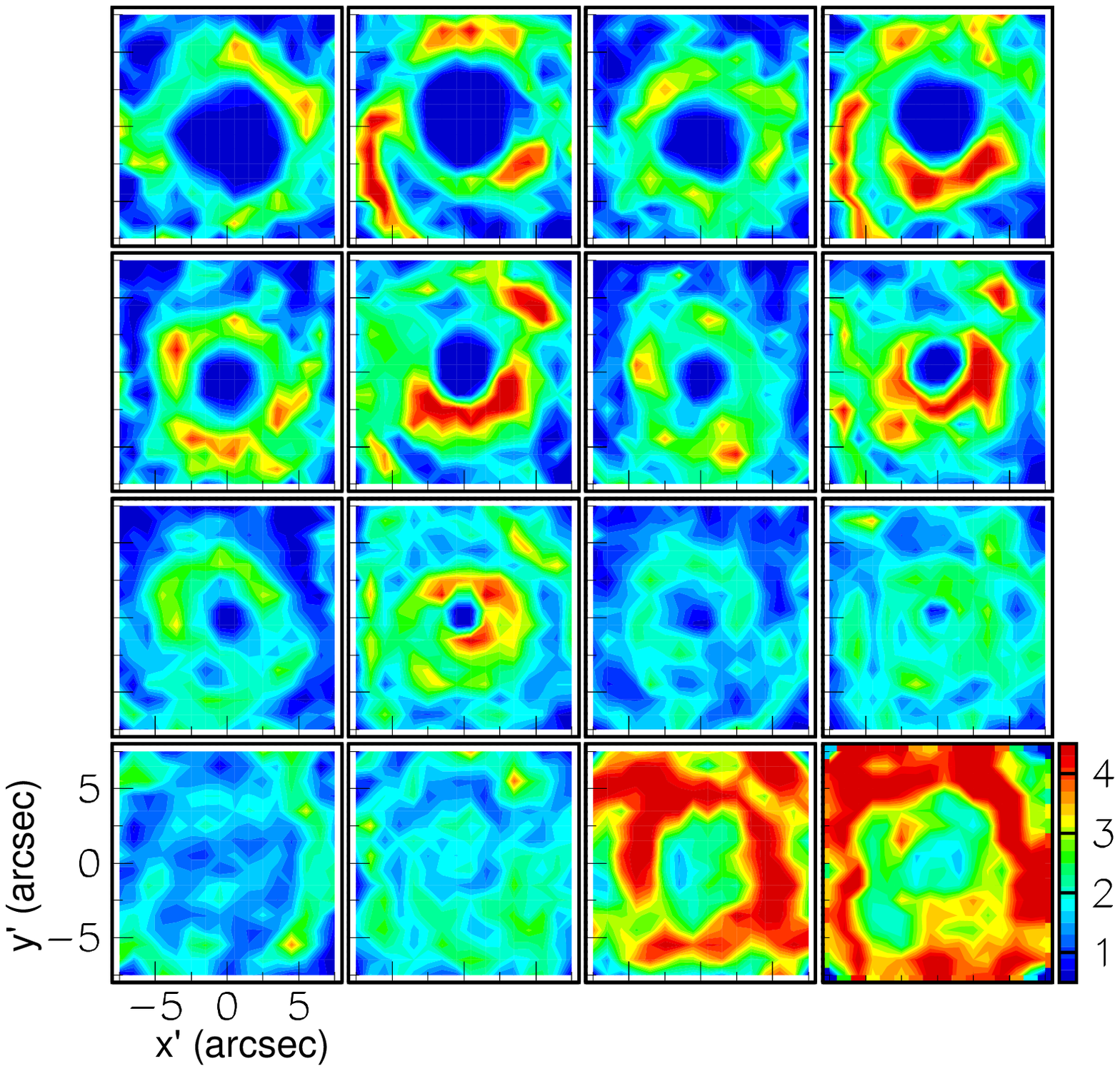}
\includegraphics[width=0.49\textwidth,trim=2cm 1.5cm 1.cm 1cm,clip]{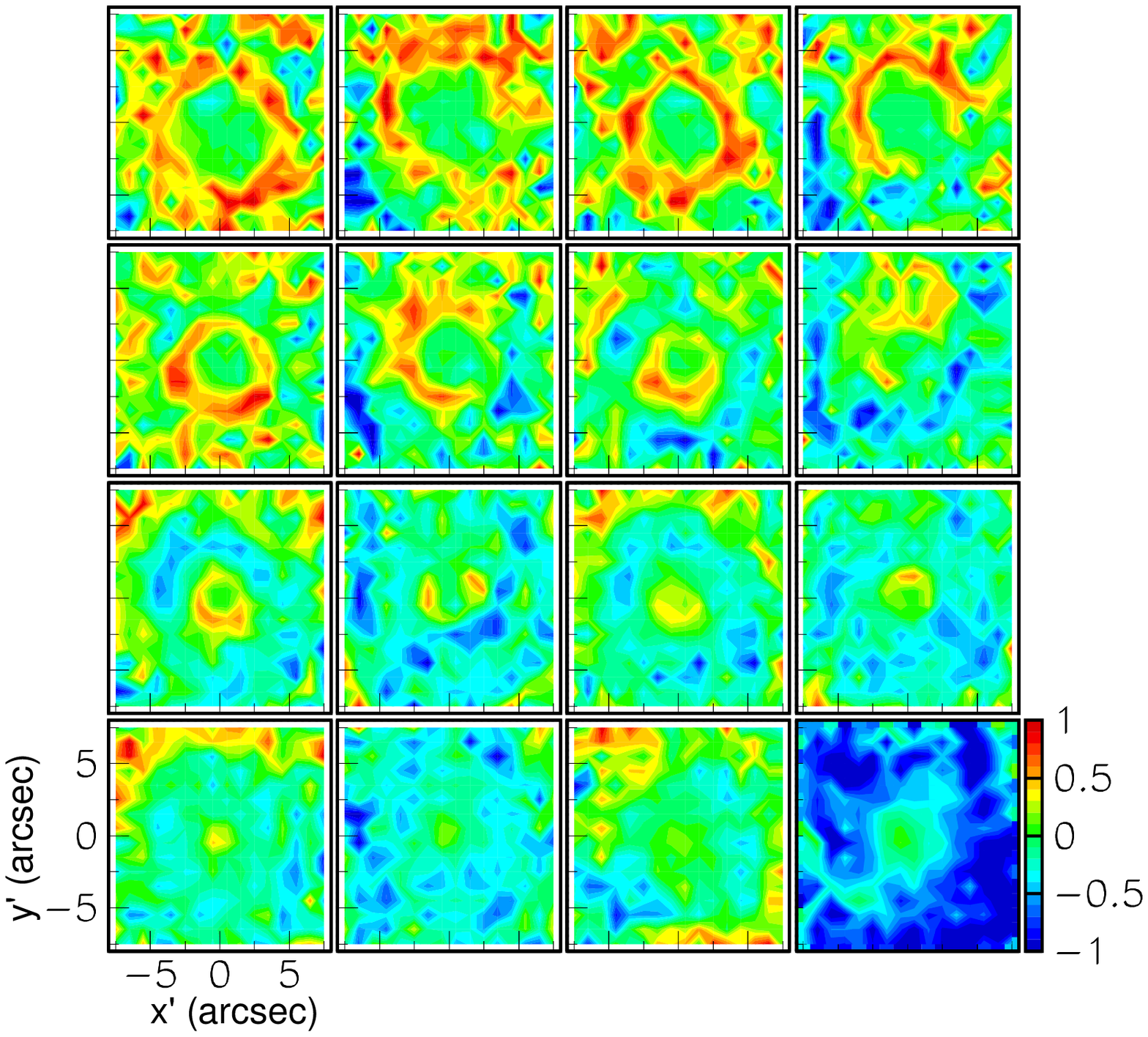}
\caption{ Maps of $\rho^*r^2$ in 1 arcsec wide $z'$ intervals for the half-sum (left)
  and half-difference (right) of \mbox{CO(1-0)} and \mbox{CO(2-1)} emissions. Maps are shown in
  pairs associated with opposite values of $z'$, negative for the leftmost, positive
  for the rightmost. From left to right and up to down, pairs are in the order of
  $z'= \pm[8,7], \pm[7,6], \pm[6,5],...,\pm[2,1], \pm[1,0]$ arcsec. The wind parameters
  are $V_{pole}=12.5$ \kms, $V_{eq}=6$ \kms, $n=1.5$, $\varphi=3.8^\circ$ and $\theta=12^\circ$.
  Units are as described in the caption of Figure \ref{fig7}.}
\label{fig11}
\end{minipage}
\end{figure*}

Finally, we recall that the irregularities of the emission that are revealed
in Figure~\ref{fig11} are in large part significant to the extent that they appear in
both \mbox{CO(1-0)} and \mbox{CO(2-1)} emissions but are of modest amplitude.

\section{Summary and conclusions}

In summary, a general analysis of recent high spatial resolution observations of the \mbox{CO(1-0)}
and \mbox{CO(2-1)} emissions of the circumstellar envelope of EP Aqr has been presented. The data
have been shown to display an intrinsic non-spherical axi-symmetry making a  small angle with respect to the line of sight, constraining, in practice, the effective
emissivity and the wind velocity distribution to obey similar axi-symmetry. Differences
between the properties of \mbox{CO(1-0)} and \mbox{CO(2-1)} emissions have been shown to be consistent
with the effect of a reasonable temperature gradient, the main point being an excellent
agreement between the two. A detailed analysis of these observations in terms of density
and temperature, rather than effective emissivity as in the present article, is the subject
of another publication (Hoai et al., in preparation), which addresses issues that have been
ignored here, in particular concerning effects of absorption and optical thickness. While
such effects do not affect the qualitative conclusions of the present analysis, they do
have significant quantitative consequences on the interpretation of the effective emissivity
in terms of temperature and density. A preliminary radiative transfer calculation shows that
absorption effects reach values between 10\% and 40\% on the effective emissivity over the meridian plane.

We have presented arguments that tend to prefer bipolar to equatorial winds, on the
basis that the latter generally require higher wind velocities in order to account for the
totality of the Doppler velocity spectra and produce a flux of matter decreasing faster with
distance from the star. Such high wind velocities tend to increase the depression of emission
in the polar regions, already important in any wind model because of the relatively smooth
decrease of the observed Doppler velocity spectra near their end points. An enhancement of
emission in the equatorial plane has been found necessary in order to produce the important
narrow central component observed in the Doppler velocity spectra everywhere on the sky plane.

Significant irregularities of the emission (significant in that they are seen
consistently in both \mbox{CO(1-0)} and \mbox{CO(2-1)} emissions) have been observed, however at a level
not exceeding typically 40\% and insufficient to provide clear evidence for important specific features.

Evidence for a clear influence of the orientation of the star axis on the behaviour of the narrow
Doppler velocity component in the sky plane has been provided using as figure of merit a quantity, $\chi^2_{axi}$,
that summarises elegantly the available information. However, the possible competition of rotation and
expansion in the equatorial disc or ring makes the interpretation difficult and beyond the scope of the present article.  

\begin{acknowledgements}
  We thank Dr Pierre Lesaffre for his interest in our work and useful comments. This paper makes use
  of the following ALMA data: 2016.1.00026.S. ALMA is a partnership of ESO (representing its member
  states), NSF (USA) and NINS (Japan), together with NRC (Canada), NSC and ASIAA (Taiwan), and KASI
  (Republic of Korea), in cooperation with the Republic of Chile. The Joint ALMA Observatory is operated
  by ESO, AUI/NRAO and NAOJ. This work was supported by the Programme National ``Physique et Chimie
  du Milieu Interstellaire'' (PCMI) of CNRS/INSU with INC/INP co-funded by CEA and CNES. The Hanoi team
  acknowledges financial support from VNSC/VAST, the NAFOSTED funding agency under grant number
  103.99-2015.39, the World Laboratory, the Odon Vallet Foundation and the Rencontres du Viet Nam.
  This research is funded by Graduate University of Science and Technology under grant number GUST.STS.DT2017-VL01.
  
\end{acknowledgements}

\end{document}